\documentclass[journal]{IEEEtran} 
\usepackage{cite}

\usepackage{latexsym}
\usepackage{verbatim, amsbsy}
\usepackage{algorithm}
\usepackage{algpseudocode}

\usepackage[dvips]{graphicx}
\usepackage{url}
\usepackage{color}
\usepackage{amsmath}                      
\usepackage{amssymb}
\usepackage{amsfonts}
\usepackage{amsthm}
\usepackage{times}
\usepackage[english]{babel}
\usepackage[dvips]{graphicx}
\usepackage[normalem]{ulem}
\usepackage{bm}
\usepackage{array}
\usepackage{tikz}
\usetikzlibrary{arrows,positioning,shapes.geometric,decorations.pathmorphing,arrows.meta}
\usetikzlibrary{decorations.pathreplacing,patterns}
\usepgflibrary{shapes.geometric}

\newcommand{\ben}{\begin{enumerate}}
\newcommand{\een}{\end{enumerate}}
\newcommand{\be}{\begin{equation}}
\newcommand{\ee}{\end{equation}}
\newcommand{\bea}{\begin{eqnarray}}
\newcommand{\eea}{\end{eqnarray}}
\newcommand{\bc}{\begin{cases}}
\newcommand{\ec}{\end{cases}}
\newcommand{\bi}{\begin{itemize}}
\newcommand{\ei}{\end{itemize}}

\newcommand{\spa}{\,\,\,\!\!}

\newcounter{mytempeqncnt}

\algnewcommand{\algorithmicgoto}{\textbf{go to}}%
\algnewcommand{\Goto}[1]{\algorithmicgoto~\ref{#1}}

\algnewcommand{\algorithmicbreak}{\textbf{break}}%
\algnewcommand{\Break}[0]{\algorithmicbreak}

 \theoremstyle{break}

 \newtheorem{lem}{Lemma}
 {
 }

\def\de{\mathrm{d}}

\newcommand{\figw}{0.8\columnwidth} 
\def\q{q} 

\begin{document}
\title{Resource Allocation and Sharing in URLLC for IoT Applications using Shareability Graphs}

\author{Federico~Librino,~\IEEEmembership{Member,~IEEE,}
        and~Paolo~Santi,~\IEEEmembership{Senior Member,~IEEE.}
\thanks{F. Librino is with the Italian National Research Council, 56124 Pisa, Italy (e-mail: federico.librino@iit.cnr.it).}
\thanks{P. Santi is with the Italian National Research Council, 56124 Pisa, Italy and with the Massachussets Institute of Technology, 02139 Cambridge, MA, USA (e-mail: paolo.santi@iit.cnr.it).} 
\thanks{Copyright (c) 20xx IEEE. Personal use of this material is permitted. However, permission to use this material for any other purposes must be obtained from the IEEE by sending a request to pubs-permissions@ieee.org.} }

\date{January 2020}
\maketitle

\begin{abstract}
The current development trend of wireless communications aims at coping with the very stringent reliability and latency requirements posed by several emerging Internet of Things (IoT) application scenarios.
Since the problem of realizing Ultra Reliable Low-Latency Communications (URLLC) is becoming more and more important, it has attracted the attention of researchers, and new efficient resource allocation algorithms are necessary.
In this paper, we consider a challenging scenario where the available spectrum might be fragmented across non-adjacent portions of the band, and channels are differently affected by interference coming from surrounding networks. Furthermore, Channel State Information (CSI) is assumed to be unavailable, thus requiring an allocation of resources based only on topology information and channel statistics. To address this challenge in a dense smart factory scenario where devices periodically transmit their data to a common receiver, we present a novel resource allocation methodology based on a graph-theoretical approach originally designed to allocate mobility resources in on-demand, shared transportation. The proposed methodology is compared with two benchmark allocation strategies, showing its ability of increasing spectral efficiency of as much as 50\% with respect to the best performing benchmark. Contrary to what happens in many resource allocation settings, this increase in spectrum efficiency does not come at the expense of fairness, which is also increased as compared to benchmark algorithms.
\end{abstract}

\section{Introduction}
The evolution of the 5G networks and emergence of IoT applications have led to a novel class of service, characterized by extremely high reliability and very short delay. So called \emph{Ultra-Reliable Low-Latency Communications (URLLC)} have been recognized as necessary, in the next future, for several critical applications, ranging from e-health to autonomous driving, from factory automation to remote surgery. Nonetheless, URLLC are extremely challenging, due to their stringent and conflicting constraints.
On one side, the necessary level of reliability might impose a packet error rate as low as $10^{-9}$; on the flip side, an unprecedented end-to-end latency is also required, leading to admitted packet delivery delays of a few milliseconds, or even less.
The packet dimension is also very small, with a size of a few hundreds of bits, depending on the application. The existing mobile communication systems are not able to meet all these requirements. For instance, the frame time of a Long Term Evolution (LTE) system is equal to 10 ms, which is above the latency constraint prescribed for several URLLC applications.
In these peculiar conditions, several commonly adopted assumptions do not hold any more. Wireless channel models exploiting the separability of the effects of path loss, fading and shadowing might not be the best option~\cite{M7}.
Furthermore, resource allocation schemes for URLLC cannot rely on the well-known and widely adopted Shannon capacity expression, since it assumes infinite block length, which is clearly not true in this scenario, and would thus overestimate the resulting reliability~\cite{M14}. Conversely, the performance limits for short packet communications (SPC) must instead be considered~\cite{M12}.

Nowadays, orthogonal frequency division multiple access (OFDMA) appears as a promising solution to cope with the challenges of URLLC.
In modern communications systems, OFDMA is indeed employed as an effective way to exploit multi-user diversity. Furthermore, its inherent flexibility in the allocation of the available resources makes it an ideal candidate to improve reliability and reduce end-to-end delay by properly leveraging the frequency dimension.

In most works, orthogonal resource allocation is prescribed, in order to avoid interference among transmitters, at least those located in the same area. However, non-orthogonal transmissions, if properly organized, can strongly enhance the overall system capacity by allowing an efficient reuse of the available resources.
Powerful signal processing schemes, like Successive Interference Cancellation (SIC), allow receivers to separate the signals coming from different sources, and their implementation have been tested in several types of networks~\cite{M18,M19,M20}. The rapid spread of ultra-dense networks, where radio resources become highly contended between devices, is now making SIC a pivotal technique to accommodate the huge spectrum demand in the wireless networks of the near future.

In this paper, we focus on an industrial scenario, like a smart-factory, where a high number (tens or even hundreds) of sensors are deployed, and communicate with a single access point. We consider a discrete automation application, where the packet transmissions are assumed to be periodic, and a semi-persistent cyclic reservation is required, as in~\cite{M5}. Under this scheme, the available resources within one cycle are to be allocated to the users in order to match both the reliability and the latency contraints.
Most of the existing works about URLLC resource allocation schemes usually leverage centralized algorithms, solving properly formulated optimization problems at a central entity, in order to improve key performance metrics like sum throughput, energy efficiency or system capacity. 
Perfect Channel State Information (CSI) is often assumed to be available and fed to the proposed algorithms.

Differently from commonly adopted assumptions, we consider a less idealized scenario. While we retain the assumption of centralized control, which is consistent with our focus on industrial scenario, we drop the common assumption that the coherence bandwidth is large enough to include all the system channels. Indeed, in many scenarios (including the one considered herein), multipath effects are particularly pronounced, leading to higher delay spreads. Furthermore, the available spectrum might be fragmented, with the utilized frequency channels spread over non-adjacent portions of the spectrum. For all these reasons, we consider in this paper that different channels can experience different fading conditions. 

Another commonly used assumption that is challenging to guarantee in dense industrial scenarios is that updated CSI for all the channels is exchanged with the access point. In contrast, in this paper we rely only on the {\em statistical knowledge} of the channel conditions, which is dictated by the network topology, assumed to be fixed and known. 

Finally, in real implementations URLLC systems are unlikely to be deployed in a completely interference free environment. In the future smart factory environments, frequencies will be often and intensively reused across space, and other transmitting devices in the surrounding buildings or areas are likely to generate interference on some or all of the available subcarriers. This in turns may make some channels preferable to others, thus further increasing the complexity of the allocation problem.

\subsection{Related Work}
Several transmission and resource allocation schemes for URLLC have been recently explored in the existing literature~\cite{M3,M4,M5}. Various proposals for a new frame strucure based on OFDMA have been suggested by the 3rd Generation Partnership Project (3GPP) standards committee~\cite{M23}, where slots shorter than 1 ms have been specified in order to meet the challenging latency constraints.

A major distinction among the existing approaches arises from the choice between utilizing a reservation phase and preferring grant-free channel access. The former option ensures that the frequency resources are orthogonally allocated to the transmitting devices, but it also requires additional time in order to perform the allocation and inform all the terminals. This makes it harder to meet the stringent delay requirements typical of URLLC.
Conversely, grant free protocols completely skip the reservation phase, at the cost of potentially generating interfering transmissions and thus impairing system reliability. Envisioned solutions to combat this drawback include the use of proactive retransmissions, which can increase the decoding probability without the need for acknowledgment packets.

A centralized scheme, which aims at maximizing the spectral efficiency, is proposed in~\cite{M1} for a real-time wireless communication-control system.
The system capacity, as a function of the available bandwidth, average Signal to Noise plus Interference Ratio (SINR), Quality of Service (QoS) parameters and delay constraint is studied in~\cite{M2}. Here, Hybrid Automatic Repeat Request (HARQ) retransmissions are also considered, and different ways of allocating the resources are compared, showing that capacity is enhanced by extending URLLC transmissions in time as much as possible while using the least amount of bandwidth.

The usage of HARQ is investigated in details in~\cite{M9}, where the problem of balancing the resources reserved to the uplink and to the downlink in URLLC is tackled.
Uplink and downlink resources are both optimized also in~\cite{M11,M13}, where multiple antennas are employed in order to exploit spatial diversity in place of temporal diversity, which is highly reduced in URLLC due to the extremely short admitted delay. Similarly, authors in~\cite{IoT} analyze the error probability of URLLC when massive multiuser Multiple Input Multiple Output (MIMO) techniques are implemented, showing its potential in effectively combating severe shadow fading and guaranteeing reliability for IoT applications.

The main drawback of centralized schemes is that the resource allocation problem in URLLC-OFDMA systems is often intractable, since the corresponding optimization problem is non convex. Suboptimal solutions, which can be implemented in polyonial time, have been pursued as well~\cite{M6, M8, M16}.
A successive convex approximation is exploited in~\cite{M15}, where URLLC are applied to Device-to-Device (D2D) communications leveraging the opportunistic spectrum sharing principle and an achievable rate maximization problem, subject to the outage probability constraint, is formulated and transformed into a tractable one. ~\cite{M6} employs tools from Artificial Intelligence (AI) in order to provide a fast hybrid resource allocation which entails also non orthogonal sharing, while 
a study about the wireless energy transfer in the URLLC scenario has been presented in~\cite{M16},
where the optimal block length for energy and information transfer is derived.

\subsection{Our contribution}
In order to perform efficient resource allocation in the considered scenario, we exploit an analogy between the problem at hand and the allocation of mobility resources in the context of online transportation -- see Section \ref{sec:genfra} for an in-length discussion. The main methodological contribution of this paper is showing how to adapt graph-theoretical approaches originally introduced in the context of on-demand mobility to the context of a prominent IoT communication challenge, namely, URLLC radio resource allocation. More specifically, we design a novel iterative algorithm that allocates the available resources taking into account sensor locations (and thus the corresponding channel statistics), as well as the interference levels of all the channels. Further leveraging the analogy with {\em shared} on-demand mobility and the concept of {\em Shareability Network} \cite{M24,M25}, we also introduce {\em channel sharing} in the context of URLLC. While orthogonal resource allocation is optimal, since it avoids cross interference among devices, it limits the amount of data which can be exchanged, especially when the network is dense and spectrum resources are limited. Channel sharing, if properly planned, can instead vastly increase the spectral efficiency.
We hence extend our proposed method by exploiting Successive Interference Cancellation (SIC) at the access point. SIC is widely considered a valid technique to implement resource sharing, and efficient scheduling schemes based on it have been proposed, with both perfect and imperfect cancellation~\cite{M17,M22,M21}.

To summarize, the original contributions of this paper are the following:
\begin{itemize}
 \item We design a novel, fast iterative algorithm to perform resource allocation in a URLLC system with only statistical CSI, for periodic packet arrival patterns;
 \item We compare the performance of this algorithm with two benchmark allocation schemes, proving its effectiveness in increasing the number of served devices;
 \item We devise a novel strategy to incorporate Successive Interference Cancellation into the proposed algorithm by exploiting the concept of shareability network.
\end{itemize}

The rest of the paper is structured as follows. In Section~\ref{sec:genfra}, we describe our general framework devised for orthogonal resource allocation in URLLC. Section~\ref{sec:sic} extends the model by describing a graph-based algorithm to implement resource sharing, based on Successive Interference Cancellation, on top of the proposed framework. In order to assess the algorithm performance, in Section~\ref{sec:sysmod} we give a detailed description of a possible application scenario, including the traffic pattern, the channel modeling and the data transmission details. Two additional benchmark allocation algorithms are then described in details in Section~\ref{sec:compallo} for comparison, while the results are shown in Section~\ref{sec:resu}. Section~\ref{sec:conclu} concludes the paper.

\section{General Framework}
\label{sec:genfra}
We consider a smart factory scenario, where $N$ devices need to periodically convey relevant information to a single central controller via radio communications. The messages are typically short, but must be reported with very high reliability. Furthermore, a stringent delay constraint is imposed, and messages which are not delivered in time are considered lost.

We assume a communication system employing OFDMA. The available bandwidth is divided into $C$ channels, each consisting of $n_c$ adjacent subcarriers. We consider a dense, resource constrained network deployment where $N\gg C$. In our general scenario, the channels do not need to be contiguous in the spectrum. Time is divided into slots of duration $\tau$ ms, corresponding to $n_t$ OFDM transmission symbols, and we define a Resource Unit (RU) as the basic time-frequency transmission resource, which spans over one channel and one time slot.
The packet generation process is deterministic and periodic at all the transmitters, with period $\nu$ ms, which corresponds to $T=\nu/\tau$ consecutive RUs and is called a \emph{cycle}.
In each cycle, every device $D_i$ generates a data packet at a fixed issue time $t_i\in[1,2,\ldots,T]$, and must deliver the packet to the access point within a delay of $\Delta_i$ slots.

Overall, as shown in Fig.~\ref{fig:cycle}, there are $CT$ available RUs per cycle, which must be efficiently allocated to the $N$ devices in order to allow them to deliver their data packets with reliability $\rho$ without violating the delay constraint.

\begin{figure}
 \centering
 \begin{tikzpicture}[>=stealth, scale = 0.6]
  \foreach \x in {0,1,...,9}
    \foreach \y in {0,1,...,4}
      \filldraw[draw=black,fill=blue!20] (\x, \y) rectangle +(1,1);
  \draw[thick,->] (-0.8,-0.8) -- (10.5,-0.8);
  \draw[thick,->] (-0.8,-0.8) -- (-0.8, 5.5);
  \node at(-1.2,5)[rotate=90] {\small Frequency};
  \node at(10,-1.1) {\small Time};
  \node at(0.5,-0.4) {1};
  \node at(1.5,-0.4) {2};
  \node at(2.5,-0.4) {...};
  \node at(9.5,-0.4) {T};
  \node at(-0.4,0.5) {1};
  \node at(-0.4,1.5) {2};
  \node[rotate=90] at(-0.4,2.5) {...};
  \node at(-0.4,4.5) {C};
  
  \foreach \x in {9,9.2,...,9.8}
    \foreach \y in {4,4.2,...,4.8}
      \filldraw[draw=black,fill=blue!20] (\x, \y) rectangle +(0.2,0.2);
  
  \draw[decorate,decoration={brace,amplitude=5pt}] (10,5) -- +(0,-1) node[black,midway,xshift=0.5cm,yshift=-0.8cm,rotate=270]{$n_c$ subcarriers};
  \draw[decorate,decoration={brace,amplitude=5pt}] (9,5) -- +(1,0) node[black,midway,yshift=0.4cm,xshift=-0.2cm]{$n_t$ symbols};
 \end{tikzpicture}
 \caption{During a cycle, up to $CT$ RUs can be allocated, where each RU consists of $n_t$ subsequent OFDM symbols spread over $n_c$ adjacent subcarriers (forming a channel).}
 \label{fig:cycle}
 \vspace{-0.5cm}
\end{figure}
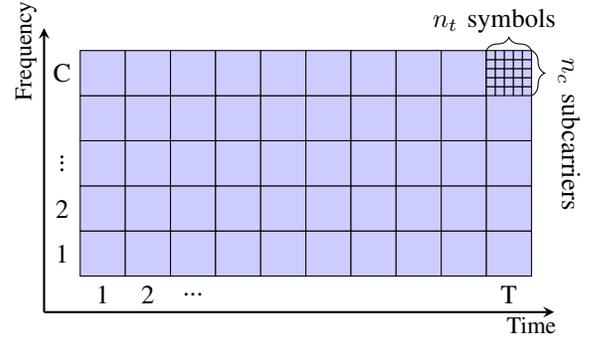

\begin{figure*}[tb]
 \centering
 \begin{tikzpicture}[>=stealth,scale = 0.9]
  \tikzset{chan/.style={draw=blue,shape=circle,fill=blue!30,minimum size=0.3cm, inner sep = 0.01cm},
           algo/.style={draw=black,shape=rectangle,fill=yellow,minimum width=1.5cm,minimum height=1cm, inner sep = 0.1cm}}
  \foreach \y in {0,...,9}
    \filldraw[draw=black,fill=blue!50] (0,\y/10*3) rectangle +(0.3,0.3);
    \node at(0,-0.5) {\small Set of devices};
    
    \filldraw[draw=black,fill=red!30] (1.2,-0.8) rectangle +(6.3,3.8);
    \draw[thick,red,->] (0.5,1.5) -- (1.5, 1.5);
    
    \node at (1.8,1.8) [chan] (dev1) {};
    \node at (1.9,2.6) [chan] (dev2) {};
    \node at (2.1,1) [chan] (dev3) {};
    \node at (2.6,0.5) [chan] (dev4) {};
    \node at (2.6,1.7) [chan] (dev5) {};
    \node at (2.9,2.5) [chan] (dev6) {};
    \node at (3.7,2.1) [chan] (dev7) {};
    \node at (3.5,1.4) [chan] (dev8) {};
    \node at (3.8,0.6) [chan] (dev9) {};
    \node at (4.4,1.1) [chan] (dev10) {};
    
    \draw[-] (dev1) to (dev2);
    \draw[-] (dev2) to (dev6);
    \draw[-] (dev5) to (dev6);
    \draw[-] (dev1) to (dev5);
    \draw[-] (dev1) to (dev3);
    \draw[-] (dev3) to (dev4);
    \draw[-] (dev5) to (dev4);
    \draw[-] (dev5) to (dev7);
    \draw[-] (dev5) to (dev8);
    \draw[-] (dev6) to (dev7);
    \draw[-] (dev8) to (dev9);
    \draw[-] (dev9) to (dev10);
    \draw[-] (dev8) to (dev10);
    \draw[-] (dev7) to (dev8);
    \draw[-] (dev4) to (dev8);
    
    \node at (3.1, -0.5) {\small Shareability Graph};
    
    \draw[->,red,thick] (4.7,1.5) -- (5.7,1.5);
    
    \foreach \y in {0,...,5}
      \filldraw[draw=black,fill=green!50] (6,\y/10*3+0.6) rectangle +(0.3,0.3);
    \node at (6.15,-0.1) {\small Set of};
    \node at (6.15,-0.5) {\small equivalent devices};
    
    \draw[->,red,thick] (6.5,1.5) -- (7.9,1.5);
    
    \node at (8.9,1.5) [algo] (sma) {GBA};
    
    \draw[->,red,thick] (9.9,1.5) -- (10.7,1.5);
    
    \begin{scope}[xshift=0.2cm]
         \foreach \x in {0,...,7}
      \foreach \y in {0,...,4}
         \draw[black] (10.8+\x/5,1+\y/5) rectangle +(0.2,0.2);
    
    \foreach \x in {0,...,3}
      \filldraw[draw=black,fill=red!80] (11+\x/5,1.8) rectangle +(0.2,0.2);
    \foreach \x in {0,...,4}
      \filldraw[draw=black,fill=blue!80] (11.4+\x/5,1.6) rectangle +(0.2,0.2);
    \foreach \x in {0,...,2}
      \filldraw[draw=black,fill=yellow!80] (10.8+\x/5,1.4) rectangle +(0.2,0.2);
    \filldraw[draw=black,fill=yellow!80] (12.2,1.4) rectangle +(0.2,0.2);
    \foreach \x in {0,...,5}
      \filldraw[draw=black,fill=green!80] (10.8+\x/5,1.2) rectangle +(0.2,0.2);
    \foreach \x in {0,...,4}
      \filldraw[draw=black,fill=purple!80] (11.4+\x/5,1) rectangle +(0.2,0.2);
    
    \node at (11.6,-0.5) {\small RUs allocation};
    \end{scope}

 \end{tikzpicture}
 \caption{Proposed approach for RU allocation with channel reuse. The highlighted part is optional, and provides the implementation of the resource sharing based on the Shareability Graph approach.}
 \label{fig:scheme}
 \vspace{-0.6cm}
\end{figure*}
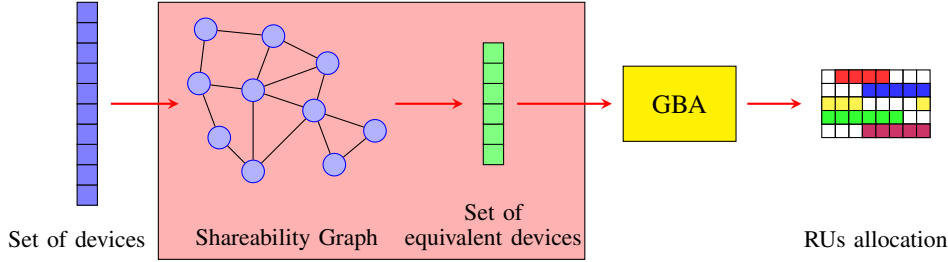

In this section, we outline a general, graph-based framework, called GBA (Graph-Based Algorithm) to perform an orthogonal resource allocation. In the next one, we define an efficient algorithm to add resource sharing through Successive Interference Cancellation on top of GBA.
This algorithm leverages the analogy between the resource allocation problem (RAP) under investigation and the vehicle dispatching problem (VDP), which has been addressed in several recent works on smart mobility systems. In both cases, a finite set of resources (frequency channels or vehicles) must be efficiently allocated to serve requests (transmissions or trips) generated by a set of users (devices or travellers) within a maximum tolerable delay.
Furthermore, resource sharing in RAP can be paralleled to vehicle sharing in VDP, and our implementation of SIC is based on the concept of \textit{Shareability Network}\footnote{While the original term used in \cite{M24,M25} to denote the data structure used to support the optimization algorithms is shareability {\em network}, we use herein the equivalent term shareability {\em graph} instead. This is to avoid confusion with the use of the term network as ``communication network" used in this paper.}  developed to efficiently solve the ride-sharing problem~\cite{M24,M25}.
A detailed discussion on the analogies between the two problems is reported in Appendix~\ref{app:analogy}.

In this paper, we envision a two-step procedure: firstly, the shareability network approach is used to determine pairs of devices which can share the same RUs. This generates a new set of \textit{equivalent devices}, some of which actually corresponding to pair of transmitters. Secondly, the GBA framework is applied to this new set, thus obtaining the desired allocation.
Our proposed approach does not depend on specific channel or decoding models, nor on the system parameters, and can hence be tailored to different instances of the problem. It is graphically sketched in Figure~\ref{fig:scheme}.

\subsection{The Graph Based Allocation Algorithm (GBA)}
The Graph Based Algorithm (GBA) starts by building a bipartite graph: the first set of vertices is the set of the $C$ channels; the second set of vertices is the set of the $N$ devices. An edge $(c,i)$ exists if device $D_i$ can transmit on channel $c$ (currently without any other assigned device) within the delay constraint. Since at the beginning all the channels are free (no users are assigned to any channel), the condition for the existence of edge $(c,i)$ is simply
\begin{equation}
 \mathcal{F}(c,i,\rho) \leq \Delta_i,
\end{equation}
where $\mathcal{F}(c,i,\rho)$ is the amount of RUs required for user $D_i$ to transmit on channel $c$ with reliability $\rho$.
If the delay constraint is not too stringent, this first bipartite graph is likely to be complete.

Since we want to allocate the highest possible amount of devices, we need to select, among the edges, those which can offer the target reliability with the lowest amount of resources. We can achieve this target by properly weighing the edges of the graph.
The weight assigned to edge $(c,i)$ is
\begin{equation}
 w(c,i) = T+\Delta_i -\left(\max(\beta_c,t_i-1) + \mathcal{F}(c,i,\rho)\right),
 \label{weights}
\end{equation}
where $\beta_c$ indicates the last busy time slot on channel $c$. At the beginning of the algorithm, all the $\beta_c$'s are set to 0. In the definition of weight $w(c,i)$, the subtracted term corresponds to the time slot at which the data transmission from user $D_i$ would be finished if it was performed on channel $c$. It takes into account both the time $t_i$ at which the data packet is generated at $D_i$ and the total transmission time $\mathcal{F}(c,i,\rho)$.
The overall weight thus corresponds to the remaining available slots on channel $c$ after allocating user $D_i$, plus an offset $\Delta_i$. This offset is added in order to avoid negative weights since, due to the periodicity of allocated slots, we can consider for allocation also slots that are \emph{wrapped around}, that is, the first slots on the same channel (to be utilized in the subsequent cycle), provided that they have not been already assigned.

Once the weighted graph has been built, the maximum weight matching can be computed in polynomial time using the Hungarian Algorithm~\cite{hungarian}. The edges included in the matching correspond to valid assignments: if edge $(c,i)$ is included in the matching, then user $D_i$ is assigned to channel $c$, meaning that the RUs $(c,t)$, with $\max(t_i-1,\beta_c)+1\leq t\leq \max(t_i-1,\beta_c)+\mathcal{F}(c,i,\rho)$ are allocated to user $D_i$.
Notice that in order to maximize the total weight, the devices with the earliest data generation times are likely to be chosen. Furthermore, the existence of the edges guarantees that the delay constraint is matched for all the assigned devices.
However, only up to $C\ll N$ devices can be allocated in this phase. Therefore, GBA then proceeds by updating the $\beta_c$'s. If edge $(c,i)$ has been included in the matching, $\beta_c$ is updated to $\max(t_i-1,\beta_c)+\mathcal{F}(c,i,\rho)$.
By doing this, we keep track of the slots on channel $c$ now assigned to user $D_i$, and the new value of $\beta_c$ still indicates the last busy slot on channel $c$ after the allocation of $D_i$.
The already allocated devices are now removed from the set of devices, and a new bipartite graph is built between the $C$ channels and the still non allocated devices.
Now, the existence condition for edge $(c,i)$ is
\begin{equation}
 \max(\beta_c,t_i-1) + \mathcal{F}(c,i,\rho) < t_i + \Delta_i,
 \label{edgecond}
\end{equation}
in order to ensure that the delay constraint is satisfied. If the vertex corresponding to a device has degree 0 in the new bipartite graph, that device is excluded from the set of transmitting devices, since it cannot be assigned to any channel.
The weights are computed again as per (\ref{weights}), and the maximum weight matching is derived, thus allocating another subset of devices. The algorithm proceeds until all the devices have been allocated or excluded by the set of transmitting devices\footnote{A minor modification is required to account for time wrapping. When checking the existence of an edge, if the transmission duration would require to occupy an already busy slot in the subsequent cycle, the algorithm does not allocate that slot, but look for free subsequent slots on the same channel, up to the packet delay limit. Hence, GBA can allocate non adjacent slots to the same device.}.
We assume that the resource allocation procedure is repeated periodically, since in real scenarios both the topology and the interference levels on the radio channels are likely to change over time, although at a slower pace. Hence, devices excluded from transmission in the current allocation round may have the opportunity to transmit in the following ones.

\begin{figure}
 \centering
 \begin{tikzpicture}[>=stealth, scale = 0.6]
  \tikzset{chan/.style={ball color=blue!30,circle,minimum size=0.7cm, inner sep = 0.01cm},
           devi/.style={circle,ball color=red!30,minimum size=0.7cm, inner sep = 0.01cm}}
           
  \node at (1.5, 4.5) {Phase 1};
  
  \node at (0, 2.4) [chan] (chan1) {$c_1$};
  \node at (0, 0) [chan] (chan2) {$c_2$};
  \node at (0, -2.4) [chan] (chan3) {$c_3$};
  
  \node at (3, 4) [devi] (devi1) {$D_1$};
  \node at (3, 2.4) [devi] (devi2) {$D_2$};
  \node at (3, 0.8) [devi] (devi3) {$D_3$};
  \node at (3, -0.8) [devi] (devi4) {$D_4$};
  \node at (3, -2.4) [devi] (devi5) {$D_5$};
  \node at (3, -4) [devi] (devi6) {$D_6$};
  
  \draw[black] (chan1) to (devi1);
  \draw[black] (chan1) to (devi2);
  \draw[black] (chan1) to (devi3);
  \draw[black] (chan1) to (devi4);
  \draw[black] (chan1) to (devi5);
  \draw[red,thick] (chan1) to (devi6);
  
  \draw[black] (chan2) to (devi1);
  \draw[red,thick] (chan2) to (devi2);
  \draw[black] (chan2) to (devi3);
  \draw[black] (chan2) to (devi4);
  \draw[black] (chan2) to (devi5);
  \draw[black] (chan2) to (devi6);
  
  \draw[black] (chan3) to (devi1);
  \draw[black] (chan3) to (devi2);
  \draw[black] (chan3) to (devi3);
  \draw[black] (chan3) to (devi4);
  \draw[red,thick] (chan3) to (devi5);
  \draw[black] (chan3) to (devi6);
  
  \draw[dashed] (4, 4.5) -- (4, -4.5);
  
  
  \node at (6.5, 4.5) {Phase 2};
  
  \node at (5, 2.4) [chan] (chan1) {$c_1$};
  \node at (5, 0) [chan] (chan2) {$c_2$};
  \node at (5, -2.4) [chan] (chan3) {$c_3$};
  
  \node at (8, 4) [devi] (devi1) {$D_1$};
  \node at (8, 0.8) [devi] (devi3) {$D_3$};
  \node at (8, -0.8) [devi] (devi4) {$D_4$};
  
  \draw[black] (chan1) to (devi1);
  \draw[black] (chan1) to (devi4);
  \draw[red,thick] (chan1) to (devi3);
  
  \draw[red,thick] (chan3) to (devi1);  
  \draw[black] (chan3) to (devi4);
  
  \draw[dashed] (9, 4.5) -- (9, -4.5);
  
  
  \node at (11.5, 4.5) {Phase 3};
  
  \node at (10, 2.4) [chan] (chan1) {$c_1$};
  \node at (10, 0) [chan] (chan2) {$c_2$};
  \node at (10, -2.4) [chan] (chan3) {$c_3$};

  \node at (13, -0.8) [devi] (devi4) {$D_4$};
  
  \draw[red,thick] (chan3) to (devi4);  
  
 \end{tikzpicture}
 \caption{Example of resource allocation using GBA. Three phases are requi\-red; in the resulting allocation, devices $D_6$ and $D_3$ transmit on channel $c_1$, $D_2$ transmits on channel $c_2$, while $D_5$, $D_1$ and $D_4$ transmit on channel $C_3$.}
 \label{fig:SMAex}
 \vspace{-0.4cm}
\end{figure}
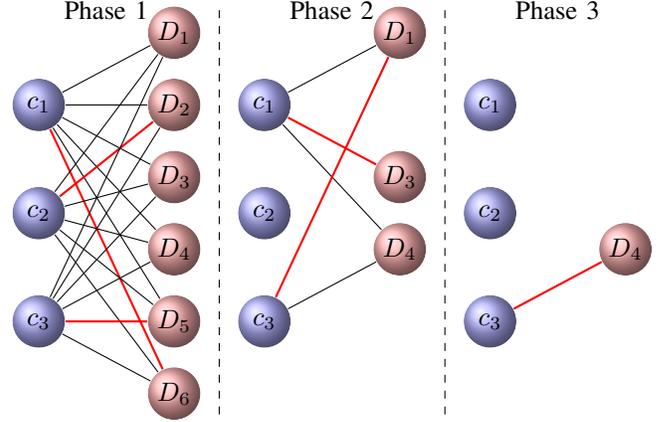

A graphic example of the implementation of GBA is illustrated in Figure~\ref{fig:SMAex}, with 3 channels and 6 devices. In the first phase, the bipartite graph is complete, and three devices ($D_2$, $D_5$ and $D_6$) are assigned to the three channels. In the second phase, they are not included in the new bipartite graph.
Notice that also some of the edges towards the remaining devices do not appear any more, due to the updated occupancy status of the channels, which may now prevent some devices to be allocated on some channels. Devices $D_1$ and $D_3$ are allocated in the second phase. The last device is finally allocated in the third phase.

The computational complexity of the GBA algorithm can be determined as follows. The algorithm proceeds by iterations, where, at each iteration, we perform the following two phases: $i)$ building/updating a bipartite graph $G$; $ii)$ computing a maximum weighted matching on $G$ using the Hungarian Algorithm. Graph $G$ has $|N|$ and $|C|$ nodes in each partition, and at most $|N|\cdot|C|$ edges. Furthermore, computing or updating the weight on each edge can be done in constant time. It follows that the computational complexity of phase $i)$ is $O(|N|\cdot|C|)$. The computational complexity of the Hungarian Algorithm when executed on bipartite graphs is $O(V^2\cdot E)$, where $V$ is the number of vertexes in the graph and $E$ is the number of edges. In our case, we have $V=|N|+|C|$ and $E=|N|\cdot|C|$, implying that the computational complexity of phase $ii)$ is $O((|N|+|C|)^2\cdot |N|\cdot |C|)$. Thus, the computational complexity of each iteration is dominated by the complexity of phase $ii)$ and equal $O((|N|+|C|)^2\cdot |N|\cdot |C|)$. Finally, we observe that at least one device is scheduled to transmit at each iteration, implying that the total number of iterations is at most $|N|$. It follows that the computational complexity of the GBA algorithm is $O((|N|+|C|)^2\cdot |N|^2\cdot |C|)$, i.e., polynomial in both $|N|$ and $|C|$.

\section{Resource Sharing based on Shareability Graphs}
\label{sec:sic}
The common assumption of exclusive RU allocation to a single transmitter, while ensuring no interference across devices, poses a limit to the number of network devices. Conversely, resource sharing, if implemented in an accurate and efficient manner, has been proved to highly improve the overall spectrum utilization.
In this section, we extend the GBA algorithm by including a Successive Interference Cancellation (SIC) mechanism at the access point, such that the same RU can be shared by two devices. In order to simplify notation and mathematical derivations, we set all the delay constraints $\Delta_i=\Delta$, but the extension to different delay bounds is straightforward.

Let us assume that a SIC procedure is implemented at the access point when two devices $D_1$ and $D_2$ are transmitting simultaneously, which works as follows:
\begin{itemize}
 \item the access point first tries to decode the signal from $D_1$, treating the signal from $D_2$ as interference;
 \item if the signal from $D_1$ has been decoded, it is subtracted from the received signal, and the access point proceeds with decoding the signal from $D_2$ without interference from $D_1$;
 \item if instead the signal from $D_1$ could not be decoded, the access point tries to decode the one from $D_2$, treating the signal from $D_1$ as interference;
 \item if the signal from $D_2$ has been decoded, it is subtracted from the received signal, and the access point tries again to decode the signal from $D_1$, now without interference from $D_2$;
 \item if neither $D_1$ nor $D_2$ can be decoded, neither signal is received and communication fails.
\end{itemize}
In order to assess when two devices can efficiently share the same resources, we need to mathematically derive the gain achievable through sharing in terms of saved resources.
Since we are applying resource sharing to the GBA algorithm, we assume in the following that the RUs allocated to the two users are all on the same channel.

\subsection{Sharing gain}
\label{sec:shagain}
\begin{figure}
 \centering
 \begin{tikzpicture}[>=Stealth, scale = 0.7]
    \def\tria{-- ++(1cm,1cm) -- ++(0cm,-1cm) -- ++(-1cm,0cm) -- cycle};
    \foreach \x in {0,1,...,9}
      \draw[draw=black] (\x, 0) rectangle +(1,1);
    
    \foreach \x in {4,5,...,8}
      \filldraw[draw=black,fill=blue!50] (\x,0) rectangle +(1,1);
    \foreach \x in {4,5,...,6}
      \filldraw[draw=black,fill=red!50] (\x,0) \tria;
    
    \draw[decorate,decoration={brace,amplitude=10pt},xshift=0pt,yshift=2pt] (4,1) -- +(5,0) node[black,midway,yshift=0.6cm]{$\mathcal{S}(c,i,j,\rho)$};
    \draw[decorate,decoration={brace,amplitude=10pt},xshift=0pt,yshift=-2pt] (7,0) -- +(-3,0) node[black,midway,yshift=-0.6cm]{$N_c(i,j)$};
    \draw[decorate,decoration={brace,amplitude=10pt},xshift=0pt,yshift=-2pt] (9,0) -- +(-2,0) node[black,midway,yshift=-0.6cm]{$K_c(i,j)$};
    
 \end{tikzpicture}
 \caption{Example of device pairing. Two users $D_i$ (red) and $D_j$ (blue) have $\mathcal{S}(c,i,j,\rho)=5$ RUs available on the same channel. However, $N_c(i,j)=3$ RUs are enough for $D_i$ to attain the target reliability, meaning that the remaining $K_c(i,j)=2$ RUs are left for $D_j$ to transmit without sharing.}
 \label{fig:equiexe1}
 \vspace{-0.4cm}
\end{figure}
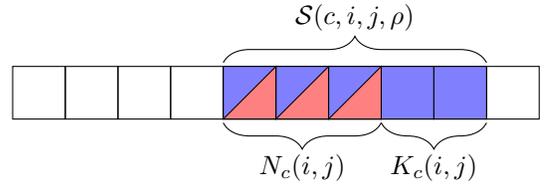

The first step in planning channel sharing is to assess when a RU can be efficiently shared between two transmitters. Let us define the number of shared RUs required for both the transmissions from devices $D_i$ and $D_j$ on channel $c$ to meet the target reliability threshold $\rho$ as $\mathcal{S}(c, i, j, \rho)$.
This function depends on the selected interference cancellation technique, as well as on the adopted channel model and decoding scheme. Due to the presence of mutual interference, we can state that
\begin{equation}
 \mathcal{S}(c,i,j,\rho) \geq \mathcal{F}(c,i,\rho), \quad\quad \mathcal{S}(c,i,j,\rho) \geq \mathcal{F}(c,j,\rho),
\end{equation}
that is, the amount of RUs required to reliably transmit the data from $D_i$ and $D_j$ on shared resources cannot be lower than the amount needed to transmit data from only one of the terminals.
Notice that in general, the two terminals $D_i$ and $D_j$ might experience different channel conditions, for instance because of their physical distance from the access point.
Therefore, the transmission from the closest device might require fewer RUs to reach the desired reliability. In this case, it is advantageous to split the assigned RUs into two subsets, yielding
\begin{equation}
 \mathcal{S}(c,i,j,\rho) = N_c(i,j) + K_c(i,j),
 \label{decompo}
\end{equation}
where $N_c(i,j)$ are the RUs which are effectively shared by both users, while $K_c(i,j)$ are those reserved only to the weaker transmitter. A graphic representation of this allocation is illustrated in Figure~\ref{fig:equiexe1}.

We define the \textit{Sharing Gain} $G_c(i,j)$ achievable by sharing the RUs between two users $D_i$ and $D_j$ on channel $c$ as the difference between the total numbers of RUs required to serve them without and with sharing:
\begin{equation}
 G_c(i,j) = \mathcal{F}(c,i,\rho)+\mathcal{F}(c,j,\rho)-(N_c(i,j)+K_c(i,j)).
\end{equation}

\subsection{Pairing the devices: the Channel Shareability Graph}
In order to decide which transmitters are to be paired, we build a graph $\mathcal{G}=(\mathcal{V}, \mathcal{E})$ called a \emph{Channel Shareability Graph}. The set $\mathcal{V}$ of vertices of the graph includes all the $N$ devices. Since the delay constraint $\Delta$ is the same for all the transmitters, each vertex $i$ of the graph, corresponding to device $D_i$, is uniquely defined by the corresponding packet issue time $t_i$.

An edge $e\in\mathcal{E}$ between two vertices $i$ and $j$ exists if the two corresponding devices $D_i$ and $D_j$ can effectively share the same resources. Hence, we need to define the \emph{shareability conditions} between a pair of devices.
Firstly, since we aim at exploiting resource sharing to {\em reduce} the number of required RUs, a condition on the gain $G_c(i,j)$ is necessary. Unfortunately, due to the chosen decoupled approach (see Fig.~\ref{fig:scheme}), the channel $c$ on which the two devices will be allocated is not known at this step yet.
Therefore, we set a stronger condition that must hold for each channel $c$, that is
\begin{equation}
 G_c(i,j) \geq 0, \quad \forall c\in\{1,2,\ldots, C\}.
 \label{condgain}
\end{equation}
This ensures that, for any possible channel on which $D_i$ and $D_j$ would possibly transmit, a non-negative gain would result from sharing the channel.

\begin{figure}
 \centering
 \begin{tikzpicture}[>=stealth,scale=0.65]
  \foreach \x in {0,1,...,9}{
    \draw[draw=black] (\x, 4.2) rectangle +(1,1);
    \draw[draw=black] (\x, 3) rectangle +(1,1);
    \draw[draw=black] (\x, 1.2) rectangle +(1,1);
    \draw[draw=black] (\x, 0) rectangle +(1,1);
  }    
  
  \foreach \x in {1,2,...,5}
    \filldraw[draw=black,fill=blue!50] (\x, 4.2) rectangle +(1,1);
  
  \foreach \x in {3,4,...,7}
    \filldraw[draw=black,fill=red!50] (\x, 3) rectangle +(1,1);
  
  \foreach \x in {0,1,...,3}
    \filldraw[draw=black,fill=yellow!40] (\x, 1.2) rectangle +(1,1);
  \filldraw[draw=black,fill=yellow!50] (9, 1.2) rectangle +(1,1);
  
  \foreach \x in {2,3,...,6}
    \filldraw[draw=black,fill=green!50] (\x, 0) rectangle +(1,1);
  
  \node at(-0.5, 4.7) {$D_1$};
  \node at(-0.5, 3.5) {$D_2$};
  \node at(11, 4.7) {$t_1=2$};
  \node at(11, 3.5) {$t_2=4$};
  
  \node at(-0.5, 1.7) {$D_3$};
  \node at(-0.5, 0.5) {$D_4$};
  \node at(11, 1.7) {$t_3=10$};
  \node at(11, 0.5) {$t_4=3$};
  
  \draw[red,thick] (3,2.8) rectangle +(3,2.6);
  \draw[red,thick] (2,-0.2) rectangle +(2,2.6);
  
  \node at(5,-1) {\small Time Slots};
  \node at(0.5,-0.4) {1};
  \node at(1.5,-0.4) {2};
  \node at(2.5,-0.4) {3};
  \node at(3.5,-0.4) {4};
  \node at(4.5,-0.4) {5};
  \node at(5.5,-0.4) {6};
  \node at(6.5,-0.4) {7};
  \node at(7.5,-0.4) {8};
  \node at(8.5,-0.4) {9};
  \node at(9.5,-0.4) {10};

 \end{tikzpicture}
 \caption{Two examples of temporal sharing conditions on channel $c$, with $T=10$ and $\Delta=5$. We assume $N_c(1,2)=3$, and $N_c(3,4)=2$. Devices $D_1$ and $D_2$ can be paired, since $\Delta-N_c(1,2)=2$ and $|t_1-t_2|=2$. Also devices $D_3$ and $D_4$ can be paired on the same channel; in fact, even if $|t_3-t_4| =7$, which is greater than $\Delta-N_c(3,4)=3$, due to the time wrapping the value $|\min(t_3,t_4)+T-\max(t_3,t_4)|$ is equal to 3, thus low enough for the two devices to be paired, according to (\ref{condtime}).}
 \label{fig:tempcond}
 \vspace{-0.4cm}
\end{figure}
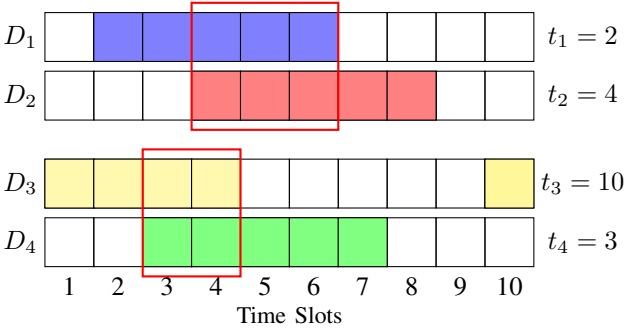

Secondly, we must ensure that the delay constraint is matched for both the devices. For any device $D_i$, with packet issue time $t_i$, call $\mathcal{T}_i$ the subset of adjacent time slots $[t_i,t_i+1,\ldots,t_i+\Delta)$. $\mathcal{T}_i$ is the \emph{transmission window} for $D_i$, meaning that all the transmissions from $D_i$ must be performed within this interval.
Now, consider a channel $c$. The number of transmissions necessary for two devices $D_i$ and $D_j$ to share the resources is given by $N_c(i,j) + K_c(i,j)$, as defined in Section~\ref{sec:shagain}. Therefore, in order for the pairing to be possible, the intersection between $\mathcal{T}_i$ and $\mathcal{T}_j$ must contain at least $N_c(i,j)$ time slots.
By turning this condition into explicit inequalities, and taking into account that time is wrapped at the end of each cycle (making it possible to potentially pair two devices transmitting at the beginning and at the end of the cycle), we get the conditions
\begin{equation}
 \min\left(|t_j-t_i|, t^+\right) \leq \Delta-N_c(i,j) \quad \forall c\in\{1,2,\ldots, C\},
 \label{condtime}
\end{equation}
with $t^+=|\min(t_i,t_j)+T - \max(t_i,t_j)|$.
Notice that the conditions above also prevent devices with a number $N_c(i,j)$ of joint transmissions higher than the maximum allowed delay $\Delta$ to be paired, which is correct. An illustrative example of the temporal sharing condition is depicted in Figure~\ref{fig:tempcond}.

However, instead of using directly (\ref{condtime}) as pairing conditions, we slightly modify them into
\begin{equation}
 \min\left(|t_j-t_i|, t^+\right) \leq \min(\Delta-N_c(i,j),M_T) \quad \forall c,
 \label{condtime2}
\end{equation}
where $M_T$ is a tunable integer parameter within the interval $[0,\Delta]$. This modification makes the conditions more stringent when $M_T<\Delta$, since it further limits the temporal distance between the packet issue times allowed for sharing. The reason behind this modification will be clarified in the next section.

When the conditions in (\ref{condgain}) and (\ref{condtime2}) are fulfilled, two devices $D_i$ and $D_j$ can be paired, and an edge between the corresponding vertices $i$ and $j$ is added to the Channel Shareability Graph $\mathcal{G}$.
The graph hence includes all the possible pairings among the devices, but only a subset of them can be actually selected. In order to maximize the number of pairings, it is sufficient to find the \emph{Maximum Matching} over the Shareability Graph, which can be done through known algorithms in polynomial time\cite{M26}.

\subsection{Generating the equivalent devices}
After identifying the pairs of devices that share resources, RU allocation can be performed. The idea is to use the GBA algorithm already described, but over a different set of devices. More specifically, we consider each pair $(D_i,D_j)$ of paired devices as an equivalent device $\hat{D}_{ij}$.
Differently from the real devices, which are fully defined by their packet issue time, $\hat D_{ij}$ requires a more detailed characterization that includes, for each of the $c$ channels, an equivalent packet issue time $\hat t_{ij}(c)$ and an equivalent allowed delivery delay $\delta_{ij}(c)\leq \Delta$.

Let us consider a couple $(D_i, D_j)$ of paired devices, and assume without loss of generality that $D_i$ is located closer to the access point. This means that, on channel $c$, $N_c(i,j)$ RUs will be required for symultaneous transmissions by $D_i$ and $D_j$, while $K_c(i,j)$ additional RUs will be necessary for the additional transmissions from $D_j$ only. As for GBA, we assume that all these RUs must be adjacent in time.
For ease of notation, in the following we use the notation $N_c$ and $K_c$ in place of $N_c(i,j)$ and $K_c(i,j)$.

We distinguish two cases: either $t_i\leq t_j$ or $t_i>t_j$. Let us analyze them both.
\begin{itemize}
 \item $t_i\leq t_j$: in this case, the deadline $t_i+\Delta$ for device $D_i$ occurs before the one for device $D_j$, thus the $N_c$ joint transmissions have higher priority than the $K_c$ additional transmissions from $D_j$, and are to be performed first.
 The earliest time slot $t_{min}(c)$ at which the joint transmissions can start is equal to $t_j$. Conversely, the latest time slot $t_{max}(c)$ at which they can start must be early enough to make it possible to allocate the $N_c$ transmissions before slot $t_i+\Delta$ and the remaining $K_c$ transmissions before $t_j+\Delta$, hence $t_{max}(c) = \min(t_i+\Delta-N_c,t_j+\Delta-N_c-K_c)$.
 \item $t_i>t_j$: in this case, it is possible to allocate some of the $K_c$ transmissions from $D_j$ before the $N_c$ joint transmissions. Since we want all the transmissions to be consecutive, the earliest time slot at which the $K_c$ transmissions can be allocated is $t_{min}(c)=\max(t_j,t_i-K_c)$.
 The latest time slot $t_{max}(c)$ at which they can start must instead guarantee that all the $N_c+K_c$ transmissions are performed before $t_j+\Delta$, hence $t_{max}(c) = t_j+\Delta-(N_c+K_c)$. Notice that since $t_i+\Delta\geq t_j+\Delta$, this also ensures that the $N_c$ joint transmissions are all performed before the deadline of $D_i$, too.
\end{itemize}
Notice that in both cases $t_{max}(c)\geq t_{min}(c)$ always holds, due to the sharing conditions (\ref{condtime2}) which are matched by $D_i$ and $D_j$.

\begin{figure}
 \centering
 \begin{tikzpicture}[>=Stealth,scale = 0.7]
    \def\tria{-- ++(1cm,1cm) -- ++(0cm,-1cm) -- ++(-1cm,0cm) -- cycle};
    \foreach \x in {0,1,...,9}
      \draw[draw=black] (\x, 0) rectangle +(1,1);
    
    \node at(1.5,2) {$t_i=2$};
    \node at(3.5,2) {$t_j=4$};
    \draw[red,thick,->] (1.5,1.8) -- (1.5,1);
    \draw[blue,thick,->] (3.5,1.8) -- (3.5,1);
    
    \node at(7,2) {Deadline for $D_i$};
    \draw[red,->] (7,1.7) -- (7,1);
    \node at(9,1.5) {Deadline for $D_j$};
    \draw[blue,->] (9,1.4) -- (9,1);
        
    \draw[purple,-] (3,-0.7) -- (3,0);
    \draw[purple,-] (8,-0.7) -- (8,0);
    
    \node at(3.5,-1) {$\hat t_{ij}$};
    \draw[purple,thick,->] (3.5,-0.8) -- (3.5, 0);
    
    \node at(5.5,-0.6) {$\color{purple}\delta_{ij}$};
    \draw[purple,<->] (3,-0.3) -- (8,-0.3);
    
    \foreach \x in {4,5,...,7}
      \filldraw[draw=black,fill=blue!50] (\x,0) rectangle +(1,1);
    \foreach \x in {4,5,...,6}
      \filldraw[draw=black,fill=red!50] (\x,0) \tria;
    
 \end{tikzpicture}
 \caption{Example of device pairing. Here $\Delta=6$, $N_c=3$, $K_c=1$ and device $D_i$ is the closest to the access point. Since its packet issue time $t_i$ is earlier than $t_j$, the earliest transmit time for the equivalent device $\hat D_{ij}$ is $t_{min} = 4$, while the last starting slot in order to accommodate all the $N_c+K_c$ transmissions is $t_{max}=5$.
 Therefore, the equivalent device has a packet issue time $\hat t_{ij}=4$ and an equivalent delay constraint $\delta_{ij}=5$. This implies that $\hat D_{ij}$ can start its transmissions on this channel only at time slot 4 or at time slot 5. This second case has been highlighted with colors, showing the 3 joint transmissions plus the additional transmission from $D_j$ only.}
 \label{fig:equiexe2}
 \vspace{-0.4cm}
\end{figure}
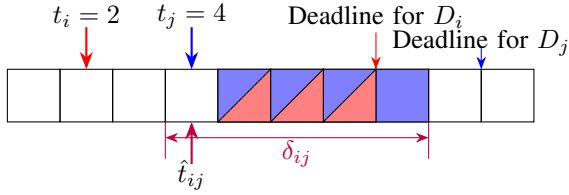

Based on the definition of $t_{min}(c)$ and $t_{max}(c)$, we can now state that the combined virtual device $\hat D_{ij}$ can be modeled as a device with packet issue time, on channel $c$, equal to $\hat t_{ij}(c) = t_{min}(c)$, and with delay constraint, on the same channel, equal to $\delta_{ij}(c) = t_{max}(c)-t_{min}(c)+N_c+K_c$.
A graphic example of device pairing is illustrated in Figure \ref{fig:equiexe2}.

The role of the limiting term $M_T$ in (\ref{condtime2}) is now clearer. As $t_i$ and $t_j$ are more distant, the intersection between the transmission windows $\mathcal{T}_i$ and $\mathcal{T}_j$ of the two devices shrinks. This in turn means that $t_{min}(c)$ and $t_{max}(c)$ become closer to each other, thus reducing the equivalent allowed delay $\delta_{ij}$.
The tuning of $M_T$ must thus be done carefully: reducing it also makes the Channel Shareability Graph sparser, but increasing it too much may prevent combined devices to be allocated due to excessively stringent delay constraints.

\subsection{GBA with Resource Sharing (GBA-SIC)}
The device pairing described in the previous section effectively defines a new, reduced set of equivalent devices to be allocated, each one now characterized by a set of $C$ packet issue times and $C$ equivalent admitted delivery delays. Notice that a device that was not paired can also be described in the same way, where its set of packet issue times are all equal to its original packet issue time, while all the equivalent delivery delays are equal to $\Delta$.

The allocation of the RUs to the equivalent devices is now performed by using the GBA algorithm. Only slight modifications are necessary, in order to take into account the presence of combined devices. Similarly to (\ref{edgecond}), the condition for the existence of an edge $(c,i+j)$ between channel $c$ and user $\hat D_{ij}$ is
\begin{equation}
 \max(\beta_c,\hat t_{ij}(c)-1) + N_c(i,j)+K_c(i,j) < \hat t_{ij}(c) + \delta_{ij}(c).
\end{equation}
Similarly, the weight assigned to the same edge is
\begin{equation}
 w(c,i+j) = T + \Delta - (\max(\beta_c,\hat t_{ij}(c)-1) + N_c(i,j)+K_c(i,j).
\end{equation}
With these adjustments, GBA can be run exactly as in the case without channel reuse, as shown in Fig.~\ref{fig:scheme}. Once the RUs have been assigned to an equivalent device $\hat D_{ij}$, it is then straightforward to distinguish the $N_c(i,j)$ ones allocated to joint transmissions from the $K_c(i,j)$ ones reserved to the additional transmissions from the farthest transmitter only.

\section{System Model}
\label{sec:sysmod}
In order to assess the performance of our resource allocation approach in a practical scenario, we need to define the functions $\mathcal{F}(c,i,\rho)$ and $\mathcal{S}(c,i,j,\rho)$. In this section, we outline a channel model suitable for URLLC, and derive the mathematical expressions accordingly.

Each transmission from device $D_i$ to the access point is subject to the combined effects of path loss and fast fading. The path loss is modeled as proportional to the distance between $D_i$ and the receiver, with exponent $\alpha$.
As to fast fading, it is assumed to be a quasi-static flat fading, and modelled as an exponential Gaussian random variable with unitary mean. Given the very short duration $\tau$ of a RU, and provided that the cycle duration $\nu$ is limited, we assume that the fading remains constant over each channel for the entire duration of a cycle, while fading coefficients across subsequent cycles and different channels are considered independent.
This last assumption represents a very general scenario, where spectrum is intensively exploited by several concurrent communication systems and the set of available channels may not be adjacent.

Coherently, we also assume that the channels are subject to weak interference coming from the other surrounding networks, which is also likely to occur in a realistic smart-factory scenario. While strong inter-networks interference can be avoided by smartly assigning the available frequencies to the different buildings/factories, weak interference may not be completely removed. We model it as additional noise, whose magnitude on channel $c$ is equal to $Y_cN_0$, being $N_0$ the noise power, while $Y_c$ is a uniform real random variable in the interval $[0,Y_M]$.
The interference magnitudes on different channels are considered to be independent.

Therefore, the SINR of $D_i$ on channel $c$ reads as
\begin{equation}
 \Gamma_{c,i} = \frac{\Gamma_T}{\Lambda_c}d_i^{-\alpha}|h_{ci}|^2,
\end{equation}
where $\Gamma_T$ is the transmit SNR, which is the same for all the transmitters, $d_i$ is the distance between $D_i$ and the access point, $h_{ci}$ is the fast fading coefficient, and $\Lambda_c=1+Y_c$.

The rate that can be used over a RU depends on the instantaneous capacity.
In URLLC, the blocklength of channel coding is too short, and the impact of decoding errors cannot be ignored. The maximal achievable rate in this scenario is thus given by \cite{M11}
\begin{equation}
 \Omega_U(c,i) \approx \frac{B}{\ln2}\left[\ln\left(1+\Gamma_{c,i}\right) - \sqrt{\frac{V}{\tau B}}Q^{-1}(\epsilon)\right],
 \label{maxrate}
\end{equation}
where $\epsilon$ is the allowed decoding error probability, $Q^{-1}(x)$ is the inverse of the Gaussian Q-function, $B$ is the bandwidth of the channel, while $V$ represents the channel dispersion, computed as
\begin{equation}
 V = 1 - \frac{1}{(1+\Gamma_{c,i})^2}.
\end{equation}

Equation (\ref{maxrate}) states that with an SINR equal to $\Gamma_{c,i}$ it is possible to transmit at rate $\Omega_U(c,i)$ with a transmission error equal to $\epsilon$, which is always greater than 0.
Equivalently, in order to transmit $b$ bits in a single RU, with a resulting rate of $b/\tau$, the error probability as a function of the instantaneous SINR is given by
\begin{equation}
 f_{err}(\Gamma_{c,i}) = Q\left(\sqrt{\frac{B\tau}{V}}\ln(1+\Gamma_{c,i}) - \frac{b\ln(2)}{\sqrt{VB\tau}}\right),
 \label{errfun}
\end{equation}
and the overall decoding error probability is obtained as
\begin{equation}
 \Psi_{c,i}(b)\! =\!\! \int_0^{+\infty}\!\!\!\!\!Q\!\left(\!\sqrt{\frac{B\tau}{V}}\ln\!\left(\!1\!+\frac{\Gamma_T}{\Lambda_c}d_i^{-\alpha}x\!\right)\! - \frac{b\ln(2)}{\sqrt{VB\tau}}\right)\!e^{-x}\!\,\de x.
 \label{errfor}
\end{equation}
The integration in (\ref{errfor}) cannot be solved in closed form. An approximation through function linearization has been proposed in~\cite{Zorzi}. Here, we instead suggest to set $V\approx1$, which is very tight when the SINR is higher than 5 dB~\cite{M12}, and approximate the shape of the error probability function $f_{err}(x)$ in (\ref{errfun}) with a step function
\begin{equation}
 f_u(\Gamma_{c,i}) =
 \begin{cases}
  1 & \text{if } \Gamma_{c,i} < \Gamma_b(\bar\epsilon) \\
  0 & \text{if } \Gamma_{c,i} \geq \Gamma_b(\bar\epsilon),
 \end{cases}
\end{equation}
where $\bar\epsilon\in(0,1)$ and $\Gamma_b(\bar\epsilon)$ is the SINR at which the target rate $b/\tau$ is achievable with error $\bar\epsilon$, which from (\ref{maxrate}) reads as
\begin{equation}
 \Gamma_b(\bar\epsilon) = \exp\left(\frac{b\ln2}{B\tau} + \frac{Q^{-1}(\bar\epsilon)}{\sqrt{B\tau}}\right) - 1.
\end{equation}

The value of $\bar\epsilon$ can be chosen arbitrarily. A threshold model is often derived by choosing it equal to 0.5. This choice yields
\begin{equation}
 \Gamma_b(\bar\epsilon) = 2^{b/(B\tau)} - 1,
 \label{appgam}
\end{equation}
which is the same threshold given by the standard Shannon capacity expression. Due to its mathematical tractability, we will adopt (\ref{appgam}) in the following, but more precise approximations, such as the linearization in~\cite{Zorzi}, can be used as well. The tightness of these approximations depends on the values of $\Gamma_T$, $\Lambda_c$ and $b$, but will be shown to be very high in the considered scenario.
The approximated decoding probability from device $D_i$ on channel $c$ with rate $b/\tau$ hence reads as
\begin{equation}
 p_{ci}(b) = \mathbb{P}[\Gamma_{c,i}\geq\Gamma_b(\bar\epsilon)] = \exp\left(-\frac{2^{b/\q} - 1}{\Gamma_Td_i^{-\alpha}/\Lambda_c}\right),
 \label{succprob}
\end{equation}
where $\q = B\tau$.

\begin{figure*}[!t]
 \normalsize
 \setcounter{mytempeqncnt}{\value{equation}}
 \setcounter{equation}{26}
 \begin{equation}
 \psi_i(\theta_i,\theta_j) =  
 \begin{cases}
 \displaystyle e^{-\lambda\theta_i-\mu\theta_j(1+\theta_i)}\! + \!\frac{\mu e^{-\lambda\theta_i}}{\lambda\theta_i+\mu}\left(\!1-e^{-s(\lambda\theta_i+\mu)}\right)\! +\! \frac{\mu}{\lambda/\theta_j+\mu}\left(e^{-s(\lambda/\theta_j+\mu)+\lambda}\! - \! e^{-\lambda\theta_i-\theta_j\mu(1+\theta_i)}\right) & \quad\text{if }\displaystyle \theta_i\theta_j<1 \\
 \displaystyle e^{-\lambda\theta_i}\left(e^{-\mu\theta_j(1+\theta_i)}\left(1 - \frac{\mu}{\lambda/\theta_j+\mu}\right) + \frac{\mu}{\lambda\theta_i+\mu}\right) & \quad\text{if }\displaystyle \theta_i\theta_j>1,
 \end{cases}
 \label{comprob}
\end{equation}
\setcounter{equation}{\value{mytempeqncnt}}
\hrulefill
\vspace{-0.3cm}
\end{figure*}

\subsection{Derivation of $\mathcal{F}(c,i,\rho)$}
Due to the extremely short duration of a RU, the transmission of a single data packet is spread over multiple RUs on the same channel. We assume that a packet is correctly received if all its parts, transmitted over different RUs, have been correctly decoded. Hence, the overall packet decoding probability depends on how the data bits are spread across the available RUs.

Let us call $b_j$ the number of bits allocated to the $j$--th available RU, such that $\sum_{j=1}^Rb_j=\ell$, where $\ell$ is the data packet length, and $R$ is the number of available RUs. In this work, we assume that any integer number of bits can be transmitted over a RU. In practical scenarios, this is not strictly true, since a transmitter has a finite set of Modulation and Coding Schemes (MCS), each one allowing to send a given number of bits over each RU. This constraint, however, can be easily incorporated in our approach, and by no means alters the validity of the proposed allocation algorithms.

The decoding probability over the $j$--th RU is given by (\ref{succprob}), with $b=b_j$. When all the RUs belong to the same channel, as prescribed in GBA, the fading coefficient, as well as the interference level, is the same over all the RUs, and the overall packet decoding probability is obtained as
\begin{equation}
 \psi_T(b_1,b_2,\ldots,b_R) = \min_j\left[\exp\left(-\frac{2^{b_j/\q}-1}{\Gamma_T d_i^{-\alpha}/\Lambda_c}\right)\right].
\end{equation}
The optimal bit allocation strategy is the one which minimizes $\max_j(b_j)$, which corresponds to equally splitting the bits across the $R$ RUs, hence $b_j = \ell/R$, $\forall j$. Under this strategy, the minimum number of RUs required in order to match the reliability constraint $\rho$ is the minimum $R^*$ such that
\begin{equation}
 \exp\left(-\frac{2^{\ell/(\q R^*)}-1}{\Gamma_T d_i^{-\alpha}/\Lambda_c}\right) > \rho.
\end{equation}
In this model, $\mathcal{F}(c,i,\rho)$ is equal to $R^*$, and is expressed as
\begin{equation}
 \mathcal{F}(c,i,\rho) = \left\lceil\frac{\ell}{\q }\left[\log_2\left(1-\frac{\Gamma_T\ln(\rho)}{\Lambda_c d_i^{\alpha}}\right)\right]^{-1}\right\rceil.
 \label{minR}
\end{equation}

\subsection{Derivation of $\mathcal{S}(c,i,j,\rho)$}
The number $\mathcal{S}(c,i,j,\rho)$ of shared RUs required for $D_i$ and $D_j$ to transmit on channel $c$ with reliability $\rho$ depends on the decoding probability of both packets.
The expression of the Shannon capacity\footnote{As in the previous section, we approximate the effective capacity with the Shannon capacity.} on channel $c$ of user $D_i$, given that user $D_j$ is also transmitting is
\begin{eqnarray}
 \Omega_S(c,i|j) & = & B\log_2\left(1 + \frac{\Gamma_T d_i^{-\alpha}|h_{ci}|^2}{\Lambda_c+\Gamma_T d_j^{-\alpha}|h_{c,j}|^2}\right) \nonumber\\
 & = & B\log_2\left(1 + \frac{\textrm{SNR}_i}{1+\textrm{SNR}_j}\right),
\end{eqnarray}
and the maximum number of bits that $D_i$ can transmit on a single RU in order for its transmission to be decoded is $\tau\Omega_S(c,i|j)$. The expression for user $D_j$ is analogous, with swapped subscripts.

The decoding probability on a RU hence depends on the number of bits that $D_i$ and $D_j$ allocate to it. According to the procedure described in the previous section, the decoding probability for device $D_i$ on channel $c$ given that $D_j$ is also transmitting is
\begin{eqnarray}
 p_{ci|j} & = & \mathbb{P}\left[\textrm{SNR}_i\geq\theta_i, \frac{\textrm{SNR}_j}{1+\textrm{SNR}_i}\geq\theta_j\right] + \nonumber \\ & + & \mathbb{P}\left[\frac{\textrm{SNR}_i}{1+\textrm{SNR}_j}\geq\theta_i, \frac{\textrm{SNR}_j}{1+\textrm{SNR}_i}<\theta_j\right],
 \label{decsic}
\end{eqnarray}
where
\begin{equation}
 \theta_i = 2^{k_i/\q}-1, \quad \quad \theta_j = 2^{k_j/\q}-1.
 \label{threij}
\end{equation}
Here, $k_i$ and $k_j$ are the number of bits sent in the considered RU by $D_i$ and $D_j$ respectively, while $\q$ is again equal to $B\tau$.
The former term in (\ref{decsic}) is the case in which the interfering user ($D_j$) is decoded first, hence allowing the interference to be removed from the signal sent by $D_i$; the latter term is instead the case in which the interfering signal cannot be decoded, and the interference must be kept while decoding the useful signal from $D_i$.

We observe that the decoding probability $p_{ci|j}$ depends on both $k_i$ and $k_j$. In the special case in which both devices send their packets over $R$ RUs, then $k_i=k_j=\ell/R$. However, as briefly discussed in Section~\ref{sec:shagain}, this is not the general case. Due to different physical locations, the two devices might need different numbers of RUs to attain the target reliability $\rho$.
In this case, some RUs are shared, while other ones are used only by one device (see Fig.~\ref{fig:equiexe1}), and the question of the optimal bit allocation across shared and non shared RUs arises.

Let us analyze the general case in which users $D_i$ and $D_j$ share a set of $R$ resources on the same channel. Assume however that $D_i$ is located closer to the access point, and hence that $R_x<R$ RUs are enough for it to complete its transmission with the target reliability, meaning that $R_y=R-R_x$ RUs are left to $D_j$ to transmit without interference.

It is clear that, for user $D_i$, all the $R_x$ RUs are equivalent, since the fading is constant, and the interference from $D_j$ is always present. Therefore, as seen in the previous section, the optimal bit allocation for $D_i$ is to spread the $\ell$ bits equally among the $R_x$ RUs.
In this work, we adopt the same strategy for $D_j$ too, thus we assume that its data bits are equally spread across all the $R$ RUs.
We highlight that this solution is clearly suboptimal, since the bits from $D_j$ are equally spread regardless of the different level of interference between the first $R_x$ RUs, where $D_i$ is also transmitting, and the remaining ones. A better solution can be attained by splitting the $\ell$ bits into two subsets, the former being then equally spread among the $R_x$ shared RUs, while the latter among the $R_y$ non shared RUs.
Finding the optimal size of the two subsets, according to the different interference level, requires solving a quite involved mathematical derivation, and for the sake of clarity is not reported in this paper but left for future work.

\addtocounter{equation}{1}
With equal bit spreading, the decoding thresholds for device $D_i$ and device $D_j$ are given by (\ref{threij}), where $k_i = \ell/R_x$ and $k_j = \ell/R$. The packet decoding probability $\psi_i(\theta_i,\theta_j)$ for user $i$ corresponds to the $p_{ci|j}$ in (\ref{decsic}), and its explicit expression is found by mathematical derivation as in (\ref{comprob}),
where $\lambda = \Lambda_cd_i^{\alpha}/\Gamma_T$, $\mu = \Lambda_cd_j^{\alpha}/\Gamma_T$ and $s = (\theta_i+1)/(1/\theta_j-\theta_i)$. The expression for $\psi_j$ is analogous, and can be obtained by swapping $\lambda$ with $\mu$ and $\theta_i$ with $\theta_j$.

Finding the exact expression of $\mathcal{S}(c,i,j,\rho)$ as a function of the decoding probability of both packets is quite involved.
Referring to equation (\ref{decompo}), $N_c(i,j)$ corresponds to $R_x$, while $K_c(i,j)$ corresponds to $R_y=R-R_x$. Hence, the problem is equivalent to find the minimum $R$ and $R_x$ such that $\psi_i(\theta_i,\theta_j)>\rho$ and $\psi_j(\theta_i,\theta_j)>\rho$.
Since an exact expression cannot be found in closed form, we rely on a heuristic algorithm.

Let us assume, without loss of generality, that $D_i$ is the closest device.
As a starting point, we set $R_x=\mathcal{F}(c,i,\rho)$ and $R=\mathcal{F}(c,j,\rho)$, that is, we start by assigning the resources which would be necessary without sharing, computed through (\ref{minR}). The (still unknown) real values cannot be lower than these ones, since the mutual interference degrades the quality of the transmissions.
From these initial values we can compute the corresponding $\theta_i$ and $\theta_j$ using (\ref{threij}), and the resulting decoding probability $\psi_i(\theta_i,\theta_j)$ for device $D_i$ using (\ref{comprob}). If this probability is lower than $\rho$, we turn one of the $R-R_x$ non shared RUs into a shared one, that is, we increase $R_x$ by 1. If it was $R=R_x$ (there were no non-shared RUs), we also increase $R$ by 1.
We repeat this until $\psi_i(\theta_i,\theta_j)\geq\rho$, thus finding the number $N_c(i,j)$ of required shared RUs.
With the obtained values of $R$ and $R_x$, we now compute $\psi_j(\theta_i,\theta_j)$. If it is lower than $\rho$, we increase $R$ by 1, that is, we add more non-shared RUs for device $D_j$. Notice that this increment also improves $\psi_i(\theta_i,\theta_j)$, since fewer bits are sent by $D_j$ over the shared RUs. We iterate this procedure until $\psi_j(\theta_i,\theta_j)\geq\rho$, thus finding also the required $R$ and hence $K_c(i,j) = R-R_x$.

\section{Benchmark Allocation algorithms}
\label{sec:compallo}
In this section, we describe two sequential resource allocation algorithms for the considered scenario, to be compared with GBA.

\subsection{Frequency Spanning Algorithm (FSA)}
The main idea of FSA is to serve each device as soon as possible. Differently from GBA, FSA can allocate to each device RUs belonging to different channels, trying to meet the reliability threshold within the smallest time interval.
The decoding probability of a data packet sent over $R$ RUs on different channels has a different expression, since different RUs have different capacities, being the fading coefficients and the interference levels independent across channels.

Let us assume, without loss of generality, that the $R$ RUs are located in channels $1,2,\ldots,V$, with $V\leq C$. Moreover, call $r_c$ the number of RUs, out of the $R$ available ones, belonging to channel $c$, such that $\sum_{c=1}^Vr_c = R$, and call $k_c$ the number of bits allocated to channel $c$. As seen in the previous section, these bits should be equally split among the $r_c$ RUs.
The overall decoding probability $\psi_c(k_c)$ of the packet fraction sent on channel $c$, according to (\ref{succprob}), hence reads as
\begin{equation}
 \psi_c(k_c) = \exp\left(-\frac{2^{\frac{k_c}{r_c}\frac{1}{\q}}-1}{\Gamma_T d^{-\alpha}/\Lambda_c}\right),
 \label{psii}
\end{equation}
where $d$ is the distance from the transmitter $D_i$ to the Access Point (AP). Correspondingly, from the independence of the fading coefficients across the channels, the overall packet decoding probability becomes
\begin{equation}
 \psi_T(k_1,k_2,\ldots,k_{V-1})\! =\! \prod_{c=1}^V\psi_{c}(k_c)\! =\! \exp\!\left(\!\!-\!\sum_{c=1}^V\frac{2^{\frac{k_c}{r_c}\frac{1}{\q}}-1}{\Gamma_T d^{-\alpha}/\Lambda_c}\!\right)\!\!,
 \label{overprob}
\end{equation}
which depends only on $V-1$ variables, since $k_V = \ell-\sum_{c=1}^{V-1}k_c$.

The bit allocation that maximizes the packet decoding probability over the selected RUs can be derived using the following Lemma:
\begin{lem}
 The optimal number of bits $k_c$ to be allocated to channel $c$, where $r_c$ RUs are available, is given by
 \begin{equation}
  k_c = \frac{\ell r_c}{R} + \frac{\q r_c}{R}\sum_{j=1}^Vr_j\log_2\left(\frac{r_c}{r_j}\frac{\Lambda_j}{\Lambda_c}\right).
 \label{optki}
 \end{equation}
 \label{lem:optki}
\end{lem}
\begin{proof}
 See Appendix~\ref{app:proflem1}.
\end{proof}
In practice, only an integer number of bits can be allocated to a RU. Hence, the optimal value given by (\ref{optki}) is then rounded to the closest integer value.

Notice that in the case of channels with the same interference level, that is, $\Lambda_c=\Lambda$ $\forall c$, and if only one RU is available per channel, that is, $r_c=1$ $\forall c$, then the optimal number of bits to be allocated in each of the $R$ RUs is simply $\ell/R$, which is reasonable, since all the RUs have the same expected capacity.
Notice also that, according to (\ref{optki}), some of the $k_c$'s might be negative (and correspondingly, other ones might be greater than $\ell$), which means that allocating any number of bits to the corresponding channels is always suboptimal. In this case, those channels should be removed from the set of the $V$ available ones, and their assigned bits split among the other ones proportionally to the number of RUs in each channel, as per the following Lemma:
\begin{lem}
If $[k_1, k_2, \ldots, k_V]$ is the optimal bit allocation among $V$ channels with a total of $R$ available RUs, then the optimal bit allocation $[\hat k_1,\hat k_2, \ldots, \hat k_{c-1}, \hat k_{c+1}, \ldots, \hat k_V]$ when the RUs on channel $c$ are removed is given by
\begin{equation}
 \hat k_j = k_j + \frac{r_j}{R-r_c}k_c, \quad \forall j\in\{1,2,\ldots,c-1,c+1,\ldots,V\},
 \label{kjmod}
\end{equation}
where $r_k$ is the number of RUs available on channel $k$.
\label{lem:modopt}
\end{lem}

\begin{proof}
 Without loss of generality, the lemma can be proved in a scenario where the last channel $V$ is removed, thus $V-1$ channels and $R-r_V$ RUs are available. Equation (\ref{optki}) in this case reads
 \begin{equation}
  \hat k_c\! =\! \frac{\ell r_c}{R-r_V} + \q r_c\left[\log_2\!\left(\frac{r_c}{\Lambda_c}\right)\! -\! \frac{1}{R-r_V}\!\sum_{j=1}^{V-1}r_j\log_2\!\left(\frac{r_j}{\Lambda_j}\right)\!\right]\!\!,
 \end{equation}
 which, through mathematical derivations, leads to (\ref{kjmod}).
\end{proof}

We observe that if we remove a channel $c$ such that $k_c<0$, this implies that in the new optimal allocation all the channels will have a reduced amount of bits allocated. This may cause other channels to have a negative amount of bits allocated, and we must proceed iteratively by removing also those channels, until all the $k_c$'s are positive.
Since the sum of the $k_c$'s is always equal to $\ell$ at each iteration, we are guaranteed that the process ends with a feasible allocation with only positive $k_c$ values.

We can now describe the functioning of FSA. The algorithm first sorts all the devices according to their packet issue times, and all the channels according to their measured interference value.
The $CT$ RUs available in a cycle are also sorted based on their location in time and frequency. We label each RU as $(c_i,s_i)$, with $1\leq c_i\leq C$ and $1\leq s_i\leq T$. The order among the RUs follows the rule
\begin{equation}
 (c_1,s_1)\prec(c_2,s_2) \Leftrightarrow (s_1<s_2) \vee \left((s_1=s_2) \wedge (\Lambda_{c_1}<\Lambda_{c_2})\right),
\end{equation}
that is, RU $(c_1,s_1)$ precedes RU $(c_2,s_2)$ if its starting time is lower, or if they start at the same time slot but channel $c_1$ has a lower interference value.

Based on this sorting, FSA scans the sorted list of devices from the first element to the last. For each device $D_i$, the algorithm allocates to it the first available RUs in the RU sorted list. The exact number of necessary RUs is derived by adding one RU at a time, computing the optimal bit distribution according to (\ref{optki}), and deriving the corresponding decoding probability through (\ref{overprob}).
Subsequent RUs are assigned to $D_i$ until its decoding probability is higher than the reliability threshold $\rho$, provided that all the RUs are within the delay constraint $t_i+\Delta$. If it is not possible to allocate enough RUs matching this condition, the device cannot be served, and is excluded by the set of transmitting devices.
Since the transmissions are organized in cycles, time can be considered to be wrapped up for the allocation purpose, and when the last devices are to be allocated, the first RUs can be checked, provided that they are free.

\begin{figure}
 \centering
 \begin{tikzpicture}[>=stealth,scale=0.7]
  \foreach \x in {0,1,...,9}
    \foreach \y in {0,1,...,3}
      \draw[draw=black] (\x, \y) rectangle +(1,1);
  \node at(-1,2)[rotate=90] {\small Sorted Channels};
  \node at(5,-0.5) {\small Time Slots};
  \node at(0.5,-0.4) {1};
  \node at(1.5,-0.4) {2};
  \node at(2.5,-0.4) {...};
  \node at(-0.4,3.5) {1};
  \node at(-0.4,2.5) {2};
  \node at(-0.4,1.5) {3};
  \node at(-0.4,0.5) {4};
  
  \filldraw[draw=black,fill=red!50] (0,3) rectangle +(1,1);
  \filldraw[draw=black,fill=red!50] (0,2) rectangle +(1,1);
  \filldraw[draw=black,fill=red!50] (0,1) rectangle +(1,1);
  
  \filldraw[draw=black,fill=blue!50] (1,3) rectangle +(1,1);
  \filldraw[draw=black,fill=blue!50] (1,2) rectangle +(1,1);
  \filldraw[draw=black,fill=blue!50] (1,1) rectangle +(1,1);
  \filldraw[draw=black,fill=blue!50] (1,0) rectangle +(1,1);
  \filldraw[draw=black,fill=blue!50] (2,3) rectangle +(1,1);
  
  \filldraw[draw=black,fill=green!50] (2,2) rectangle +(1,1);
  \filldraw[draw=black,fill=green!50] (2,1) rectangle +(1,1);
  \filldraw[draw=black,fill=green!50] (2,0) rectangle +(1,1);
  \filldraw[draw=black,fill=green!50] (3,3) rectangle +(1,1);
  \filldraw[draw=black,fill=green!50] (3,2) rectangle +(1,1);
  \filldraw[draw=black,fill=green!50] (3,1) rectangle +(1,1);
  
  \filldraw[draw=black,fill=yellow!50] (3,0) rectangle +(1,1);
  \filldraw[draw=black,fill=yellow!50] (4,3) rectangle +(1,1);
  \filldraw[draw=black,fill=yellow!50] (4,2) rectangle +(1,1);
  \filldraw[draw=black,fill=yellow!50] (4,1) rectangle +(1,1);
  \filldraw[draw=black,fill=yellow!50] (4,0) rectangle +(1,1);
  \filldraw[draw=black,fill=yellow!50] (5,3) rectangle +(1,1);
  
  \filldraw[draw=black,fill=purple!50] (6,3) rectangle +(1,1);
  \filldraw[draw=black,fill=purple!50] (6,2) rectangle +(1,1);
  
  \filldraw[draw=black,fill=red!50] (0.3,4.5) rectangle +(0.4,0.4);
  \filldraw[draw=black,fill=green!50] (1.3,4.3) rectangle +(0.4,0.4);
  \filldraw[draw=black,fill=blue!50] (1.3,4.7) rectangle +(0.4,0.4);
  \filldraw[draw=black,fill=yellow!50] (2.3,4.5) rectangle +(0.4,0.4);
  \filldraw[draw=black,fill=purple!50] (6.3,4.5) rectangle +(0.4,0.4);
  
  \node[anchor = north] (data) at(7,6.1) {\small Data Generation};
  
  \draw[->] (6,5.5) -- (3, 4.8);
  \draw[->] (7,5.5) -- (6.5, 5);
  
  \draw[->,thick] (0.5,3.5) -- (0.5,0.5);
  \draw[->,thick] (0.5,0.5) -- (1.5,3.5);
  \draw[->,thick] (1.5,3.5) -- (1.5,0.5);
  \draw[->,thick] (1.5,0.5) -- (2.5,3.5);
  
  \draw[red,thick] (0,5.5) -- (4,5.5);
  \draw[red,thick] (0,5.3) -- (0,5.7);
  \draw[red,thick] (4,5.3) -- (4,5.7);
  
  \node[anchor = north] at (2,6.1) {\small Latency constraint $\Delta$};

 \end{tikzpicture}
 \caption{An example of resource allocation for 5 devices using FSA, with 4 available channels. The channels are sorted from the best (1) to the worst (4), and the packet generation time slot of each device is highlighted on the top.}
 \label{fig:FSAex}
 \vspace{-0.5cm}
\end{figure}
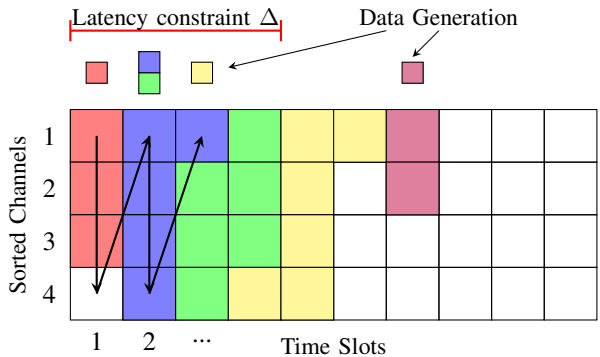

An example of allocation obtained with FSA is illustrated in Figure~\ref{fig:FSAex}. As can be observed, when network traffic is high, some retransmissions are allocated also to channels with relatively poor quality, which in turns might increase the number of necessary RUs and potentially lead to the violation of the delay constraint.
As a matter of fact, while FSA is often envisioned as an effective way of allocating resources, its effectiveness holds only when the channels have the same quality and statistics. If this is the case, it is equivalent to serialize the transmissions over the same channel in subsequent time slots or to parallelize them in the same time slot across different channels.
If instead, as in our scenario, there are ``good'' and ``bad'' channels, the overall number of required RUs to deliver a packet with the target reliability also depends on which channels are selected for transmission.
For comparison, we report in Appendix~\ref{app:corcha} the FSA optimal bit allocation and performance in the case of highly correlated channels. There, we show that channel correlation is beneficial for FSA, although the different channel qualities, represented by the different interference levels, still limit its effectiveness.

\subsection{Best Channel Algorithm (BCA)}
This algorithm is similar to GBA, but operates in a sequential manner. BCA first sorts the devices based on their packet generation time (as FSA), but subsequently allocates to each device only RUs belonging to the best available channel.
In this case, the \emph{best channel} is the one that makes the data transmission as quick as possible, in order to leave the subsequent resources available for the other devices.
Let us assume that we need to allocate device $D_i$, whose packet generation slot is $t_i$. Let us call $\beta_c$ the last slot during which channel $c$ is busy, according to the devices allocated up to now, with $0\leq\beta_c\leq T$.
The best channel $c^*$ for device $D_i$ is defined as
\begin{equation}
 c^* = \arg\min_{1\leq c\leq C}\left[\max\left(t_i,\beta_c+1\right) + \mathcal{F}(c,i,\rho)\right],
 \label{bestcha}
\end{equation}
where we use the expression for $\mathcal{F}(c,i,\rho)$ in (\ref{minR}).
Therefore, $c^*$ is the channel that minimizes the time within which the transmission can be performed, taking into account both the current occupancy status of the channel and its quality.
Once channel $c^*$ has been identified, the following condition is checked:
\begin{equation}
 \max\left(t_i,\beta_{c^*}+1\right) + \mathcal{F}(c^*,i,\rho) < t_i+\Delta.
 \label{condok}
\end{equation}
If the condition holds, the allocation is feasible, and the algorithm assigns the RUs $(c^*,j)$, with $\max\left(t_i,\beta_{c^*}+1\right)\leq j\leq \max\left(t_i,\beta_{c^*}+1\right) + \mathcal{F}(c^*,i,\rho)-1$ to device $D_i$. If instead (\ref{condok}) is false, the same holds for all the other channels as well: the device $D_i$ cannot be allocated to any channel without violating the delay contraint, and it is excluded from the set of transmitting devices.

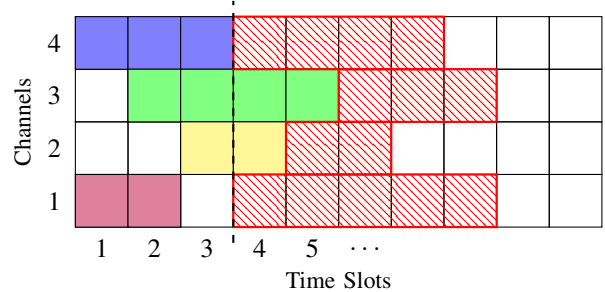
\begin{figure}
 \centering
 \begin{tikzpicture}[>=stealth, scale=0.7]
  \foreach \x in {0,1,...,9}
    \foreach \y in {0,1,...,3}
      \draw[draw=black] (\x, \y) rectangle +(1,1);
  \node at(-1,2)[rotate=90] {\small Channels};
  \node at(5,-1) {\small Time Slots};
  \node at(0.5,-0.4) {1};
  \node at(1.5,-0.4) {2};
  \node at(2.5,-0.4) {3};
  \node at(3.5,-0.4) {4};
  \node at(4.5,-0.4) {5};
  \node at(5.5,-0.4) {$\ldots$};

  \node at(-0.4,3.5) {4};
  \node at(-0.4,2.5) {3};
  \node at(-0.4,1.5) {2};
  \node at(-0.4,0.5) {1};
  
  \filldraw[draw=black,fill=purple!50] (0,0) rectangle +(1,1);
  \filldraw[draw=black,fill=purple!50] (1,0) rectangle +(1,1);
  
  \filldraw[draw=black,fill=blue!50] (0,3) rectangle +(1,1);
  \filldraw[draw=black,fill=blue!50] (1,3) rectangle +(1,1);
  \filldraw[draw=black,fill=blue!50] (2,3) rectangle +(1,1);
  
  \filldraw[draw=black,fill=green!50] (1,2) rectangle +(1,1);
  \filldraw[draw=black,fill=green!50] (2,2) rectangle +(1,1);
  \filldraw[draw=black,fill=green!50] (3,2) rectangle +(1,1);
  \filldraw[draw=black,fill=green!50] (4,2) rectangle +(1,1);
  
  \filldraw[draw=black,fill=yellow!50] (2,1) rectangle +(1,1);
  \filldraw[draw=black,fill=yellow!50] (3,1) rectangle +(1,1);

  \draw[color=red,thick,pattern=north west lines, pattern color = red] (3,3) rectangle +(4,1);
  \draw[color=red,thick,pattern=north west lines, pattern color = red] (5,2) rectangle +(3,1);
  \draw[color=red,thick,pattern=north west lines, pattern color = red] (4,1) rectangle +(2,1);
  \draw[color=red,thick,pattern=north west lines, pattern color = red] (3,0) rectangle +(5,1);
  
  \draw[dashed,thick] (3, -0.3) -- (3, 4.3);
  
 \end{tikzpicture}
 \caption{An example of resource allocation for a device generating its packet at time slot 4, using BCA. The RU $(1,3)$ on channel 1 cannot be used, since the packet generation time is 4. Even if channel 4 becomes free earlier (at time slot 4) than channel 2, channel 2 is still preferable, since it grants the lowest completion time.}
 \label{fig:BCAex}
 \vspace{-0.5cm}
\end{figure}

Similarly to FSA, minor modifications can be applied to take time wrapping into account.
In Figure~\ref{fig:BCAex}, an example of the BCA allocation is illustrated. A new device is to be allocated, and the quality of each channel determines the number of slots required on that channel. BCA then selects the channel that offers the earliest transmission completion times (channel 2 in the example), though it is not the first to become free.

\section{Results}
\label{sec:resu}
In this section, we compare the performance of the proposed algorithms over a network of $N$ devices, connected to a single access point. The devices are randomly deployed in a circular area with radius $L$, and the AP is located at its center.
The topology, which is known to the AP, is then considered fixed over all the simulation time.
Please notice that the proposed scheme is fully general, and its effectiveness does not rely on any specific parameters setup or reference technology. Nevertheless, in order to assess its performance in a realistic scenario, we tuned the relevant quantities as follows.
As in~\cite{M5,Rev1}, the subcarriers spacing is set to $\omega=15$ kHz, and each OFDM symbol spreads over $n_c=12$ adjacent subcarriers, which form a channel. The duration of an OFDM symbol is approximately $71.4\spa \rm{\mu s}$. A Resource Unit (RU) is a sequence of $n_t=2$ consecutive OFDM symbols, and therefore spans over $B=180$ kHz and lasts for $\tau=0.144$ ms.
We consider a cycle of 10 ms and a latency constraint of 5 ms, corresponding to $T=70$ and $\Delta=35$ slots, respectively. The parameters are summarized in Table~\ref{tab:para}, and all the reported results are averaged over 100 random placements.

\begin{table}
 \centering
 \caption{System Parameters}
 \begin{tabular}{l|c|c}
  \hline
  Parameter & Symbol & Value \\
  \hline
  Subcarrier spacing & $\omega$ & 15 kHz\\
  Slot duration & $\tau$ & 0.144 ms \\
  Subchannel bandwidth & $B$ & 180 kHz \\
  Transmit SNR & $\Gamma_T$ & 100 dB \\
  Path loss exponent & $\alpha$ & 3 \\
  Max interference factor & $Y_M$ & 4 \\
  Packet size & $\ell$ & 100 bit \\
  Max pairing delay & $M_D$ & 15 slot \\
  Cycle duration & $T$ & 70 slot \\
  Latency contraint & $\Delta$ & 35 slot \\
  Target reliability & $\rho$ & 0.99999 \\
  Deployment area radius & $L$ & 50 m\\
  \hline
 \end{tabular}
 \label{tab:para}
 \vspace{-0.3cm}
\end{table}

\begin{figure}
 \begin{center}    
 \includegraphics[width=\figw]{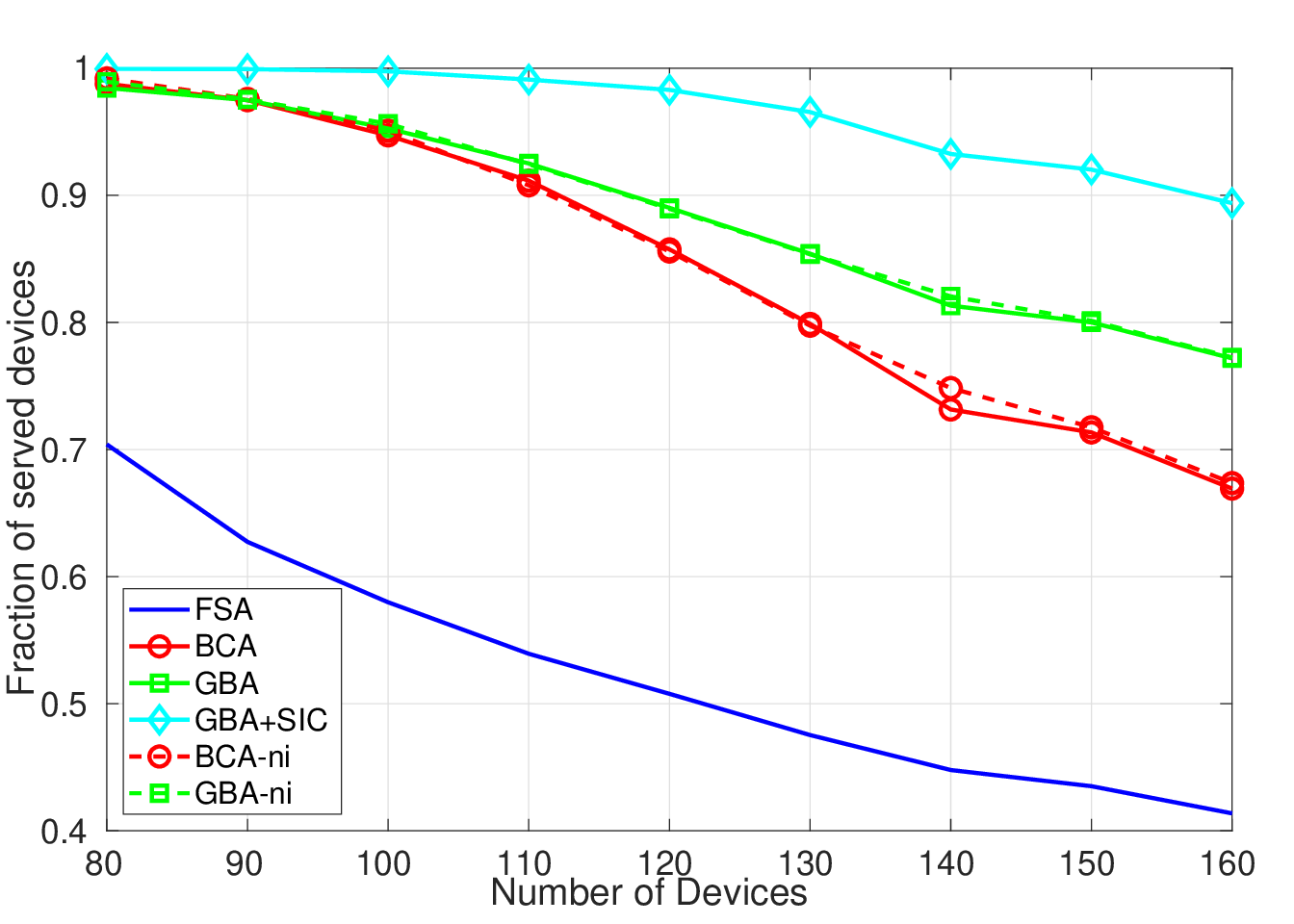}
 \caption{\small Fraction of served devices as a function of the total number of devices, with $C=7$ channels.}
 \label{fig:NEW_Fra_vs_N}
 \vspace{-0.6cm}
 \end{center}
\end{figure}

In Figure~\ref{fig:NEW_Fra_vs_N}, we set the number $C$ of available channels to 7, and plot the fraction of admitted devices (that is, devices which can transmit with the desired reliability) as a function of the total number $N$ of transmitters.
For comparison, we also plot the results for BCA and GBA when $\mathcal{F}(c,i,\rho)$ is computed directly through (\ref{errfor}) by numeric integration (\emph{-ni} suffix), which confirms the validity of our threshold approximation.
As can be immediately noticed, the strategy of allocating the earliest available resources, as implemented by FSA, is dramatically outperformed by the other two approaches. The reason lies in the fact that FSA allocates to each users RUs belonging to the available channels. In doing so, it also allocated RUs on bad channels, which can support very low transmission rates, hence hugely increasing the amount of necessary RUs.
Conversely, BCA always aims at allocating the RUs on the best available channel, even if this implies transmitting at a later time.
In this case, the limiting factor is the fact that the best resources are assigned to the devices with the earliest data packet generation times, which can be suboptimal. The GBA algorithm provides an approach with a wider perspective, where resources are assigned taking into account a global benefit (thanks to the maximum matching algorithm), and is hence capable of including more transmitters, especially when the number of devices is high.
This motivates our idea of building SIC-based resource sharing on top of GBA. The smart device pairing, obtained through the concept of channel shareability graph, allows us to share the RUs in a very advantageous manner, thus vastly increasing the system capacity. While the maximum number of devices in order for BCA and GBA to serve 95\% of the transmitters is around 100, with GBA+SIC this number is increased by more than 50\%. Looking at it in another way, when 160 devices are deployed, GBA allows successfully serving 13\% more devices than BCA; when combined with SIC, the number of served devices is further increased by more than 30\%.

\begin{figure}
 \begin{center}    
 \includegraphics[width=\figw]{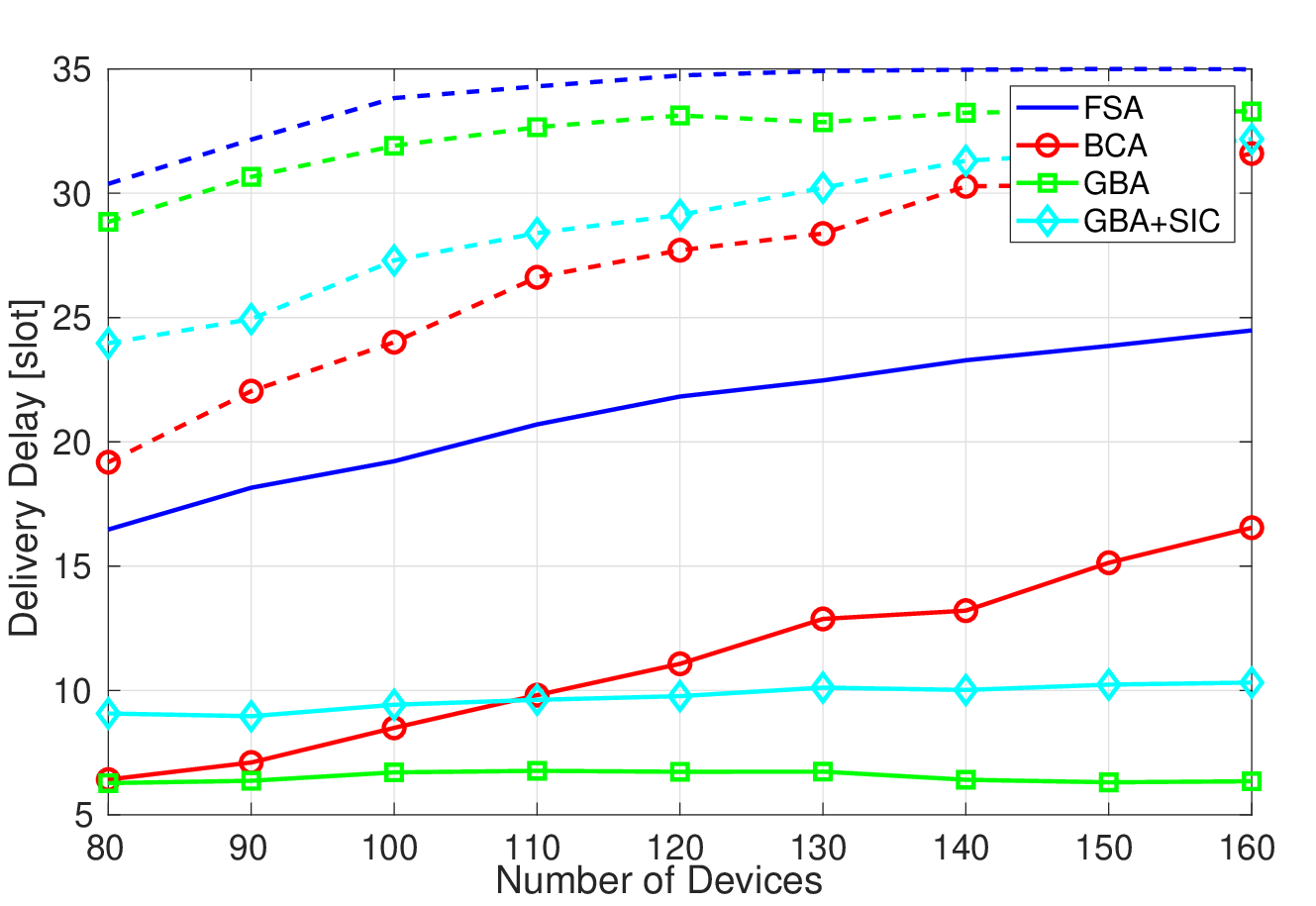}
 \caption{\small Delivery delay as a function of the number of devices. Solid lines are average values, while dashed lines are maximum values.}
 \label{fig:REA_AoI_vs_N}
 \vspace{-0.6cm}
 \end{center}
\end{figure}
While serving more users, GBA is also able to reduce the average packet delivery delay, which in turns determines the average Age of Information (AoI) at the AP. At each instant $t$, and for each transmitter $D_i$, the instantaneous AoI relative to $D_i$ is defined as the difference between $t$ and the time when the last successfully delivered packet from $D_i$ was generated.
Due to the periodic resource allocation, a new packet is received from $D_i$ every $T$ time slots\footnote{Ignoring transmission failures, which occur with probability $1-\rho\ll1$.}, with a delay $\sigma_i\leq\Delta_i$ which depends on the resource allocation. Therefore, the average AoI relative to user $D_i$ is $\sigma_i+T/2$, and the overall AoI, averaged over all the users, is obtained by adding $T/2$ to the average delivery delay.
We report in Figure~\ref{fig:REA_AoI_vs_N} the average and the maximum delivery delay for the four proposed algorithms. All the maximum values are within the maximum allowed delay $\Delta=35$ slots, which confirms the allocation validity. However, the average delay is very different among the various schemes. It quickly increases with $N$ for FSA and, although more slowly, also for BCA.
These algorithms allocate users sequentially, hence when they allocate a device over a bad channel, that channel remains busy for a long time, thus increasing the delay for all the subsequent users. Conversely, GBA iteratively allocates users such that the minimum amount of RUs is required, thus preferring devices with better channel conditions.
While a slight increase of the delay is observed when SIC is added, due to the extra time needed for allowing an efficient pairing of the users, GBA+SIC outperforms BCA in dense networks ($N\geq110$).

\begin{figure}
 \begin{center}    
 \includegraphics[width=\figw]{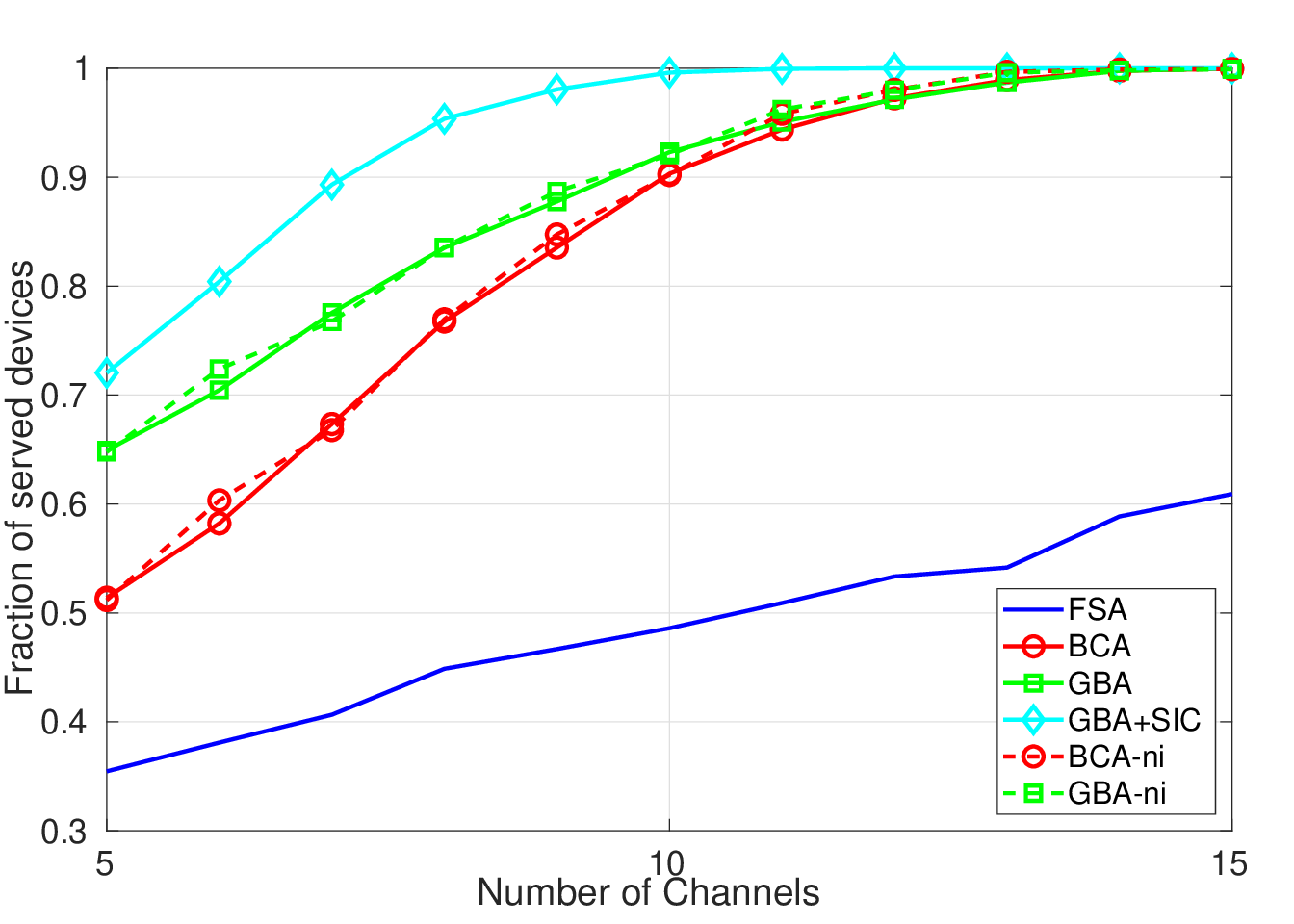}
 \caption{\small Fraction of served devices as a function of number $C$ of channels, with $N=160$ devices.}
 \label{fig:NEW_Fra_vs_C}
 \vspace{-0.6cm}
 \end{center}
\end{figure}
The effectiveness in spectrum utilization can be assessed by looking at Figure~\ref{fig:NEW_Fra_vs_C}, where we set the network cardinality to 160 and vary the number $C$ of available channels from 5 to 15. The choice of preferring the good channels instead of faster transmissions over all the available channels is confirmed by the superior performance of BCA over FSA.
When the spectrum resource is scarce, GBA can further increase the number of allocated users, thanks to its smarter channel allocation. A huge improvement is attained by employing SIC over GBA. The figure highlights the fact that only 10 channels are enough for GBA+SIC to serve all the devices, while 14 are necessary without SIC, with a considerable gain of $\sim30\%$.

\begin{figure}
 \begin{center}    
 \includegraphics[width=\figw]{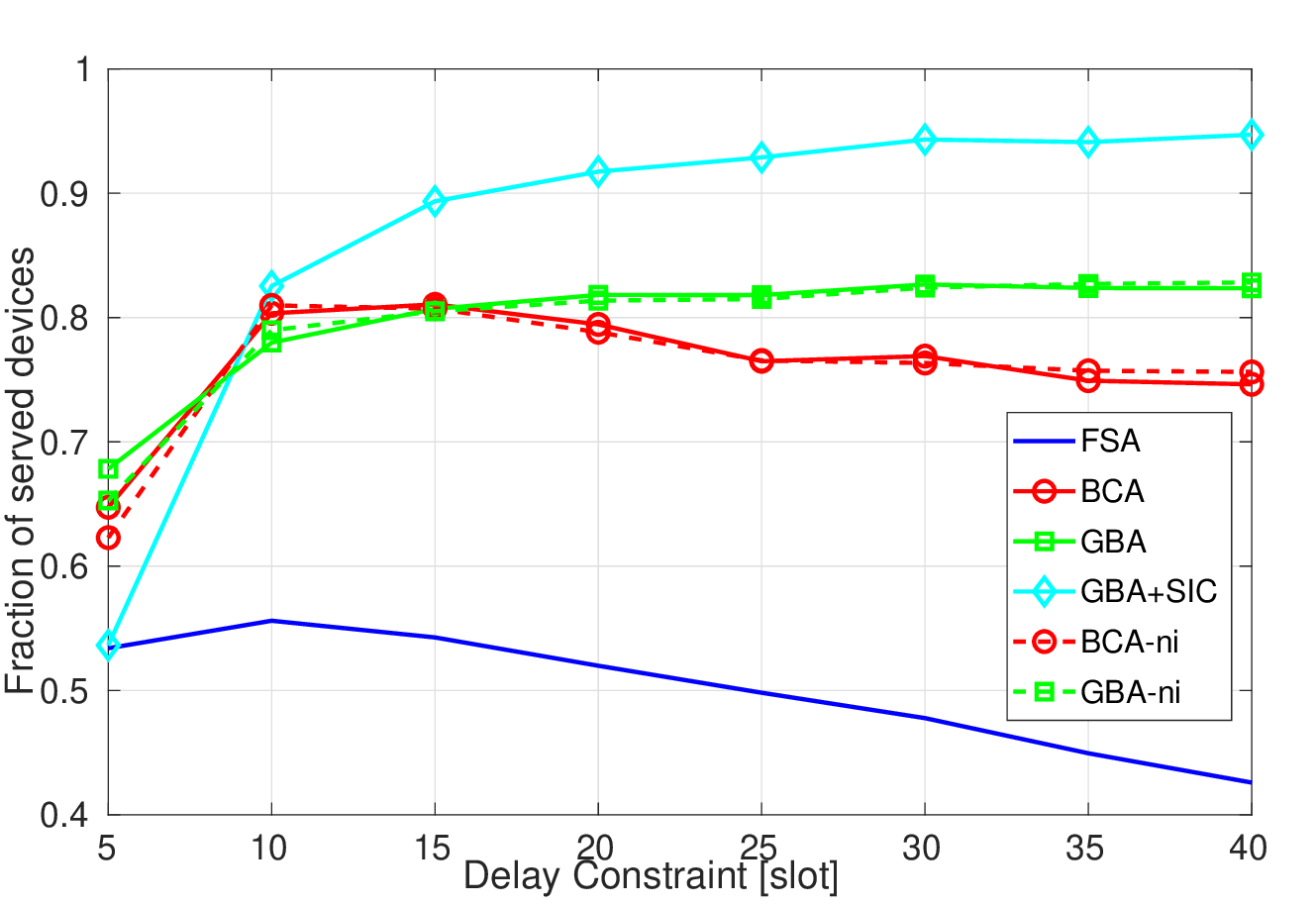}
 \caption{\small Fraction of served devices as a function of the allowed delay, with $N=140$ devices and $C=7$ channels.}
 \label{fig:NEW_Fra_vs_D}
 \vspace{-0.3cm}
 \end{center}
\end{figure}
The transmission delay constraint also plays a fundamental role in a URLLC system. We investigate this aspect in Figure~\ref{fig:NEW_Fra_vs_D}, where we vary the maximum allowed transmission delay $\Delta$. All the algorithms perform poorly when $\Delta$ is comparable to or even lower than most packets transmission time, thus preventing their transmission.
Under these conditions, it is necessary to allocate transmissions immediately after the data packet generation, and the sequential approach of FSA and BCA is not much different from the one of GBA. However, as $\Delta$ increases, GBA improves its performance, while that of FSA and BCA slowly decreases.
This impairment is due to the fact that, as the delay constraint becomes looser, even packets from the farthest transmitters, which require a high amount of RUs, are admitted, and subtract resources for the transmissions from closer devices, which could be allocated in higher number.
Conversely, there is more room for GBA to better allocate the users, holding off the transmissions which can be completed on good channels in lower time, thus granting access to a higher number of devices.
 
\begin{figure}
 \begin{center}    
 \includegraphics[width=\figw]{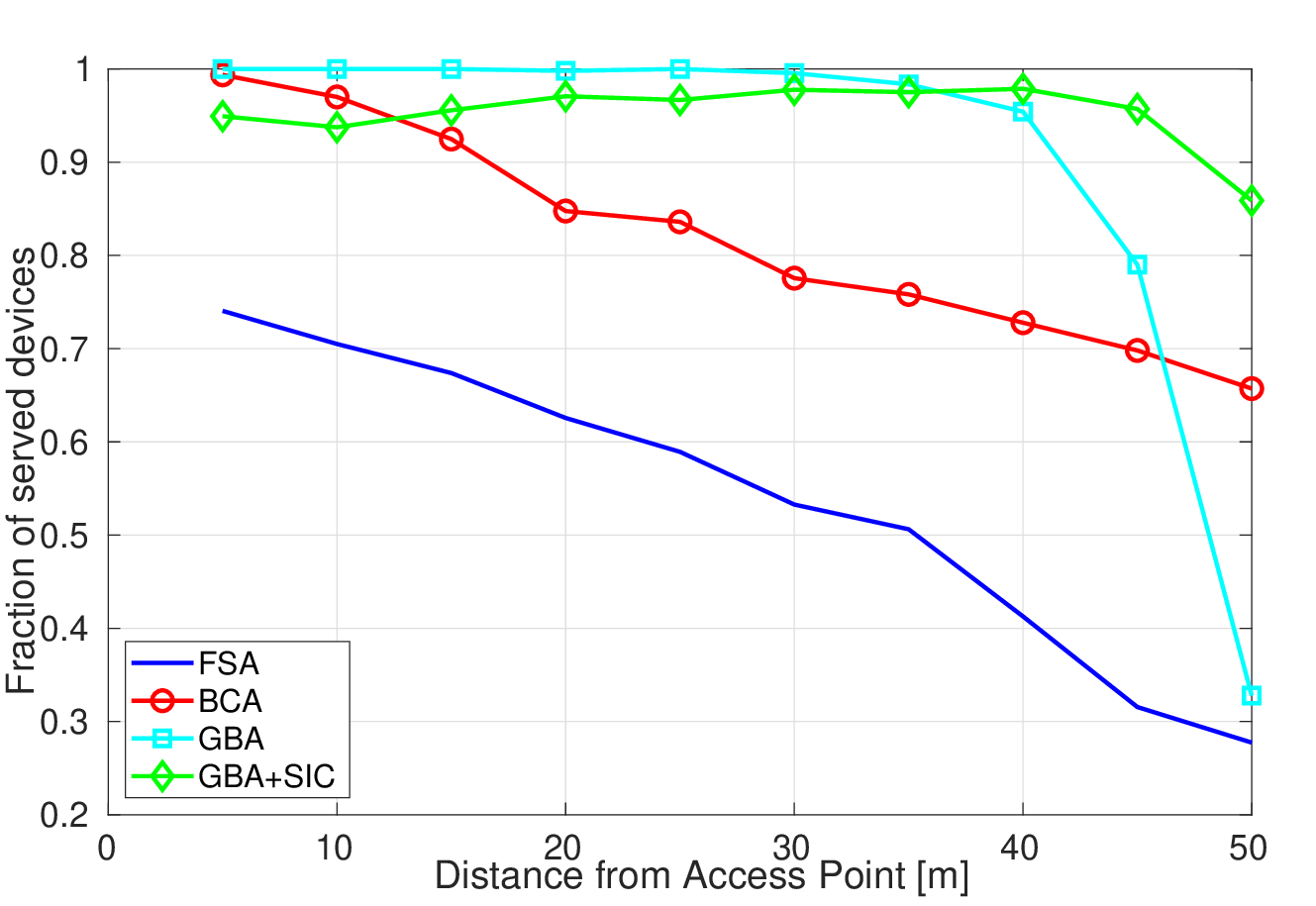}
 \caption{\small Fraction of served devices as a function of their distance from the Access Point, with $N=140$ devices and $C=7$ channels.}
 \label{fig:NEW_Fra_vs_dist}
 \vspace{-0.3cm}
 \end{center}
\end{figure}
This last observation raises an interesting point about the fairness of the proposed algorithms. Devices located farther from the AP are more resource demanding, and are likely to be excluded in order to make room for a higher number of closer transmitters. This aspect is analyzed in Figure~\ref{fig:NEW_Fra_vs_dist}, where we plot the fraction of admitted devices as a function of their distance from the AP.
In this figure, $N=140$ and $C=7$.
We can measure this metric through the Jain's fairness index
\begin{equation}
 \mathcal{F}(x_1,x_2,\ldots,x_N) = \frac{\left(\sum_{i=1}^Nx_i\right)^2}{n\sum_{i=1}^Nx_i^2},
\end{equation}
whose value for each of the four algorithms is reported in Table~\ref{tab:frafair}.
\begin{table}
 \centering
 \caption{Algorithms Fairness}
 \begin{tabular}{l|c|c}
 \hline
  \textbf{Algorithm} & \textbf{Fraction of Served Users} & \textbf{Jain's Fairness Index} \\
  \hline
  FSA & 0.4525 & 0.9258 \\
  BCA & 0.7570 & 0.9824 \\
  GBA & 0.8274 & 0.9526 \\
  GBA+SIC & 0.9474 & 0.9987\\
  \hline
 \end{tabular}
\label{tab:frafair}
\vspace{-0.7cm}
\end{table}

As expected, FSA can serve only a very limited fraction of the far users, due to its inefficient RUs allocation, achieving the lowest fairness index. BCA performs better also in terms of fairness, but its behavior is still unbalanced towards the closer transmitters, which are more easily admitted. The GBA algorithm allocates the channels taking into account both the interference levels and the topology. In doing so, it is capable of allocating a much higher number of users across almost all the network area.
In fact, the fraction of admitted devices remains constant up to the 80\% of the area radius. This, however, comes at the cost of sudden drop of this fraction towards the edge of the network, where it becomes much lower than that granted by BCA, and the resulting fairness index is also lower.
Resource sharing, introduced by GBA+SIC, can instead offer the highest fairness, with Jain's index equal to 0.9985. This is a highly notable result, as increases in spectral efficiency typically comes at the expense of fairness. Instead, GBA+SIC attains a fraction of admitted users which is only slightly lower than that of simple GBA in the central area of the network, but then keeps the same value across almost the entire network area. This is due to the fact that close and far users are often paired together. While this may prevent some close devices to be admitted, since they are paired with a very far user which requires several RUs, it also frees several RUs, which can be used for the farthest transmitters.
More refined SIC implementations, which generate equivalent devices corresponding to 3 or more real transmitters, or which jointly perform users pairing and RUs allocation, are likely to furher boost the observed improvements, and are left as a promising direction for future work.

\section{Conclusions}
\label{sec:conclu}
The main contribution of this paper is showing a specific example of successful cross-pollination across IoT-related disciplines -- on-demand mobility, which will be one of the the main applications of autonomous vehicles, and wireless communication in dense smart-factory environments.
The study presented herein demonstrates that novel methods developed to solve the problem in one setting -- on-demand mobility, can be successfully applied in another setting -- URLLC. 
A novel iterative allocation scheme, called GBA, has then been detailed for periodic traffic patterns in URLLC when CSI is not available. The proposed graph-based algorithm allows a fast resource allocation even over dense networks, and takes into account channel diversity in terms of interference level, thus better balancing the resources among devices.
Furthermore, a resource sharing scheme based on Successive Interference Cancellation has been added to the proposed allocation algorithm. The obtained results show that the combination of GBA and SIC can vastly improve spectrum utilization, by either increasing the number of admitted transmitters or reducing the amount of spectrum needed. In addition, the resulting algorithm is proved to reduce the average Age of Information and to attain a higher fairness across users, thus allowing a more balanced traffic exchange.

While providing a first example of the effectiveness of graph-based resource allocation methods in the context of URLLC, this study opens several directions for future work, such as considering application of the methodology in settings where the devices are mobile, or their topology is time-varying. Also, the issue of longer-term fairness in channel access -- i.e., the fact that a certain device will eventually get access to a fair share of the radio resources across multiple resource allocation periods -- is another interesting avenue for further research.

\appendices

\section{Proof of Lemma~\ref{lem:optki}}
\label{app:proflem1}
In order to find the optimal bit allocation, the derivatives of the decoding probability (\ref{overprob}) with respect to all the $V-1$ variables must be put to zero. For the $i$--th variable, the derivative is
\begin{eqnarray}
 \frac{\textrm{d}\psi_T}{\textrm{d}k_i} & = & \exp\left(-\sum_{i=1}^V\frac{2^{\frac{k_i}{r_i}\frac{1}{Q}}-1}{\Gamma d^{-\alpha}\Lambda_i}\right)\times \nonumber \\
 & \times & \left(\frac{\Lambda_Vd^{\alpha}\ln2}{\Gamma q r_V}2^{\frac{\ell-\sum_jk_j}{r_V\q}} - \frac{\Lambda_id^{\alpha}\ln2}{\Gamma \q r_i}2^{\frac{k_i}{r_i\q}}\right),
\end{eqnarray}
which becomes zero if
\begin{equation}
 \frac{\Lambda_i}{r_i}2^{\frac{k_i}{\q}} = \frac{\Lambda_V}{r_V}2^{\frac{\ell-\sum_{j=1}^{V-1}k_j}{\q}},
\end{equation}
and thus yields
\begin{equation}
 k_i = \ell\frac{r_i}{r_V} + \q r_i\log_2\left(\frac{r_i}{r_V}\frac{\Lambda_V}{\Lambda_i}\right) - \frac{r_i}{r_V}\sum_{j=1}^{V-1}k_j.
  \label{equadev}
\end{equation}
Here, $k_i$ still appears on both sides of the equation. Due to the symmetry of the channels, equation (\ref{equadev}) must hold for all the $V-1$ variables. By summing all the corresponding equations, we get
\begin{eqnarray}
 \sum_{i=1}^{V-1}k_i &\hspace{-0.3cm} = &\hspace{-0.3cm} \frac{\ell}{r_V}\sum_{i=1}^{V-1}\!r_i\! + \q\!\sum_{i=1}^{V-1}\!r_i\log_2\!\left(\!\frac{r_i}{r_V}\frac{\Lambda_V}{\Lambda_i}\!\right)\! -\! \frac{1}{r_V}\!\sum_{i=1}^{V-1}\!r_i\!\sum_{j=1}^{V-1}\!k_j \nonumber\\
 &\hspace{-1.8cm} = &\hspace{-0.9cm} \frac{\ell}{r_V}(R-r_V) + \q\sum_{i=1}^{V-1}r_i\log_2\!\left(\!\frac{r_i}{r_V}\frac{\Lambda_V}{\Lambda_i}\!\right)\! -\! \frac{R-r_V}{r_V}\sum_{j=1}^{V-1}\!k_j \nonumber \\
 &\hspace{-1.8cm} = &\hspace{-0.9cm} \ell\left(1-\frac{r_V}{R}\right) + \frac{\q r_V}{R}\sum_{i=1}^{V-1}r_i\log_2\left(\frac{r_i}{r_V}\frac{\Lambda_V}{\Lambda_i}\right),
 \label{totasum}
\end{eqnarray}
where we used the fact that $\sum_{i=1}^{V-1}r_i = R-r_V$.
Now, the expression of the summation obtained in (\ref{totasum}) can be inserted in the right hand side term of (\ref{equadev}), thus getting
\begin{eqnarray}
 k_i &\hspace{-0.2cm} = &\hspace{-0.2cm} \q r_i\log_2\left(\frac{r_i}{r_V}\frac{\Lambda_V}{\Lambda_i}\right)\! + \frac{\ell r_i}{R} - \frac{\q r_i}{R}\sum_{j=1}^{V-1}r_j\log_2\left(\frac{r_j}{r_V}\frac{\Lambda_V}{\Lambda_j}\right) \nonumber \\
 &\hspace{-0.2cm} = &\hspace{-0.2cm} \frac{\ell r_i}{R} + \frac{\q r_i}{R}\left[R\log_2\left(\frac{r_i}{r_V}\frac{\Lambda_V}{\Lambda_i}\right) + \right. \nonumber \\
 & & \left. -\log_2\left(\frac{\Lambda_V}{r_V}\right)(R-r_V) - \sum_{j=1}^{V-1}r_j\log_2\left(\frac{r_j}{\Lambda_j}\right)\right] \nonumber\\
 &\hspace{-0.2cm} = &\hspace{-0.2cm} \frac{\ell r_i}{R} + \frac{\q r_i}{R}\left[R\log_2\left(\frac{r_i}{\Lambda_i}\right) - \sum_{j=1}^Vr_j\log_2\left(\frac{r_j}{\Lambda_j}\right)\right],
\end{eqnarray}
which finally leads to (\ref{optki}).

\section{Analogy with Smart Mobility Systems}
\label{app:analogy}
The resource allocation problem (RAP) considered in this paper shows relevant analogies with a well known resource allocation problem in the context of transportation. In the vehicle dispatching problem (VDP), we have a number of vehicles (e.g., cabs, or a set of UBER/Lyft vehicles) that must be dispatched to serve a set of trip requests, each characterized by an origin and destination. As a fundamental Quality of Service parameter in demand-responsive mobility is the time needed for a vehicle in the fleet to pickup a passenger -- this is fundamental to reduce trip cancellations, the goal in VDP is to maximize the number of trip requests that can be served within a certain maximum pickup delay. 
The analogy between RAP and VDP then becomes clear: in RAP, resources are wireless channels, and service requests are generated by a set of transmitting devices, each with a maximum service delay to be satisfied. Similarly, in VDP resources are vehicles, and service requests are generated by travelers, each with a maximum tolerable pickup delay. 
Furthering the analogy, in VDP the serving time of a traveller depends on the specific trip to be performed, as well as on an exogenous factor: traffic conditions. In RAP, the amount of time slots reserved to a transmitter $D_i$ on channel $c$ can be modeled as a function $\mathcal{F}(c, i, \rho)$, where $\rho$ is the target reliability. The exogenous factor in this case is the interference that is experienced on each channel, which is incorporated into the definition of $\mathcal{F}(c, i, \rho)$. The exact definition of this function depends on the specific channel and decoding model, but any function can be used in our proposed framework.

There are also differences between the two settings: the switching time between two subsequent tasks (that is, serving two different transmitting devices) is zero in RAP, while it is equal to the travel time needed to reach the next traveler in case of VDP. However, for the purpose of resource optimization this does not impair the analogy.

The analogy between VDP and RAP also extends to include resource sharing. The most recent research directions for smart mobility aim at developing sustainable solutions involving ride-sharing or car-pooling, in which multiple travelers share the same vehicle. Similarly, two or more transmitting devices can share the same subset of RUs by means of SIC. 
The problem of finding good sharing solutions in VDP, while taking into account the delay constraints, is extremely challenging, especially when the number of vehicles scales up and the solution needs to be computed in real time. Traditional approaches based on variants of the traveling salesman problem cannot be used for that purpose. Nonetheless, the novel approach devised in~\cite{M24,M25} leverages the concept of \textit{Shareability Network} to identify which trip pairings offer the maximum advantage in terms of saved time or mileage. In this work, we exploit the same approach and apply it to solve the non-orthogonal RAP.

\section{Correlated Channels Scenario}
\label{app:corcha}
In this Appendix, we shed light on the impact of correlation across channels on the performance of the proposed algorithms. We consider a scenario where fading coefficients over different channels are still modeled as complex Gaussian random variables of zero mean and unitary variance, but they are no longer independent across channels. We instead retain the assumption of independence across subsequent cycles.

We first observe that, since no CSI is available at the transmitters, all the algorithms still rely only on stochastic information. The statistics of the fading of a single channel are unaltered, and the allocation and long run performance of BCA and GBA do not change.
Conversely, FSA distributes the data bits according to the joint statistics of the channels, and is hence affected by correlation.
The mathematical derivation of the optimal bit allocation over correlated channels for FSA is not straightforward. Therefore, we focus on a scenario with the highest possible correlation, that is, where channel fading in each cycle is the same across all the channels.

Let us assume that $R$ RUs are available over $V$ channels, with $V\leq C$, and call $r_c$ the number of RUs available on channel $c$, such that $\sum_{c=1}^Vr_c = R$. Call $k_c$ the number of bits assigned to the RUs on channel $c$, which are equally split across the $r_c$ RUs. The decoding probability on channel $c$ is still given by (\ref{psii}).
Since the fading $h$ at each cycle is constant across the channels, now the overall packet decoding probability is not given by (\ref{overprob}), but is equal to the decoding probability over the worst channels, that is
\begin{eqnarray}
 \psi_T(k_1,k_2,\ldots,k_{V-1}) & = & \mathbb{P}\left[|h|^2>\max_c\left(\frac{2^{\frac{k_c}{r_c}\frac{1}{\q}}-1}{\Gamma_Td^{-\alpha}/\Lambda_c}\right)\right] \nonumber \\
 & = & \!\!\!\min_c\left[\exp\left(-\frac{2^{\frac{k_c}{r_c}\frac{1}{\q}}-1}{\Gamma_Td^{-\alpha}/\Lambda_c}\right)\right]\!,
 \label{totdec}
\end{eqnarray}
which depends on $V-1$ variables, since $k_V = \ell-\sum_{c=1}^{V-1}k_c$. The optimal bit allocation is the one which maximizes $\psi_T$, and since the exponential functions in (\ref{totdec}) are non negative and monotonically increasing, it corresponds to the one which makes the decoding probability equal over all the channels, hence
\begin{equation}
 \Lambda_i(2^{\frac{k_i}{r_i\q}}-1) = \Lambda_j(2^{\frac{k_j}{r_j\q}}-1), \quad \forall i,j\in\{1,2,\ldots,V\}.
\end{equation}
Through mathematical manipulation, we get
\begin{equation}
 k_i = r_i\q\log_2\left(\frac{\Lambda_j}{\Lambda_i}\left(2^{\frac{k_j}{r_j\q}}-1\right)+1\right),
 \label{relaeq}
\end{equation}
and without loss of generality, we can express all the $k_i$'s as a function of $k_1$ through this equality (including $k_1$ itself). Now, since the data packet has size $\ell$, we can write
\begin{equation}
 \sum_{i=1}^Vr_i\q\log_2\left(\frac{\Lambda_1}{\Lambda_i}\left(2^{\frac{k_1}{r_1\q}}-1\right)+1\right) = \frac{\ell}{\q}.
 \label{equatot}
\end{equation}
Equation (\ref{equatot}) cannot be solved in closed form, but a solution for $k_1$ can be easily found numerically using the bisection method, since $0\leq k_1\leq\ell$. The obtained value $k_1^*$ is enough to get the overall decoding probability, which with this optimal bit allocation simply reads as
\begin{equation}
 \psi_T(k_1^*) = \exp\left(-\frac{2^{\frac{k_1^*}{r_1}\frac{1}{\q}}-1}{\Gamma_T d^{-\alpha}/\Lambda_1}\right),
\end{equation}
while the other $k_i$'s can be derived through (\ref{relaeq}).

\begin{figure}
 \begin{center}    
 \includegraphics[width=\figw]{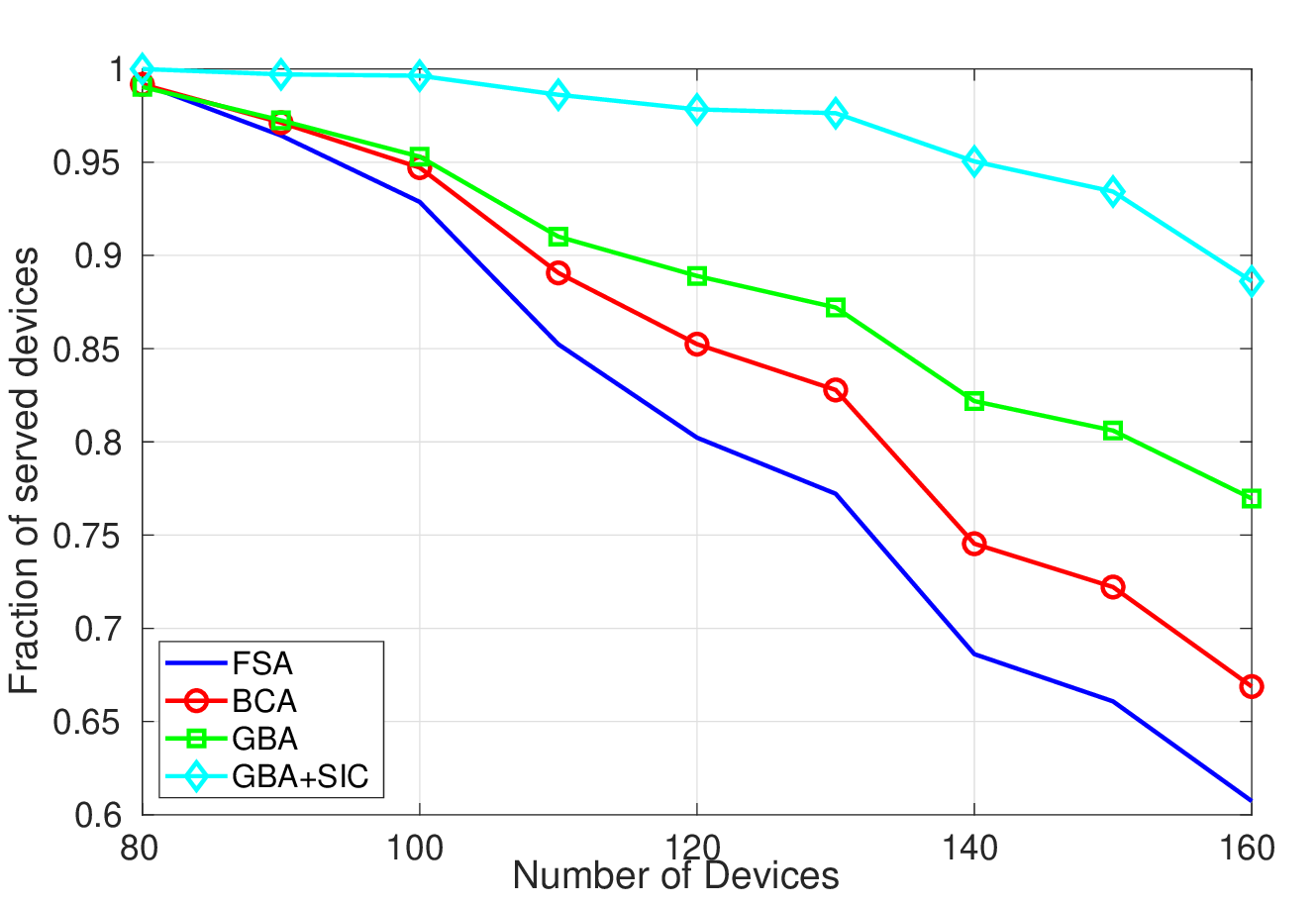}
 \caption{\small Fraction of served devices as a function of the total number of devices, with $C=7$ channels.}
 \label{fig:NEW_CON_vs_N}
 \vspace{-0.6cm}
 \end{center}
\end{figure}
\begin{figure}
 \begin{center}    
 \includegraphics[width=\figw]{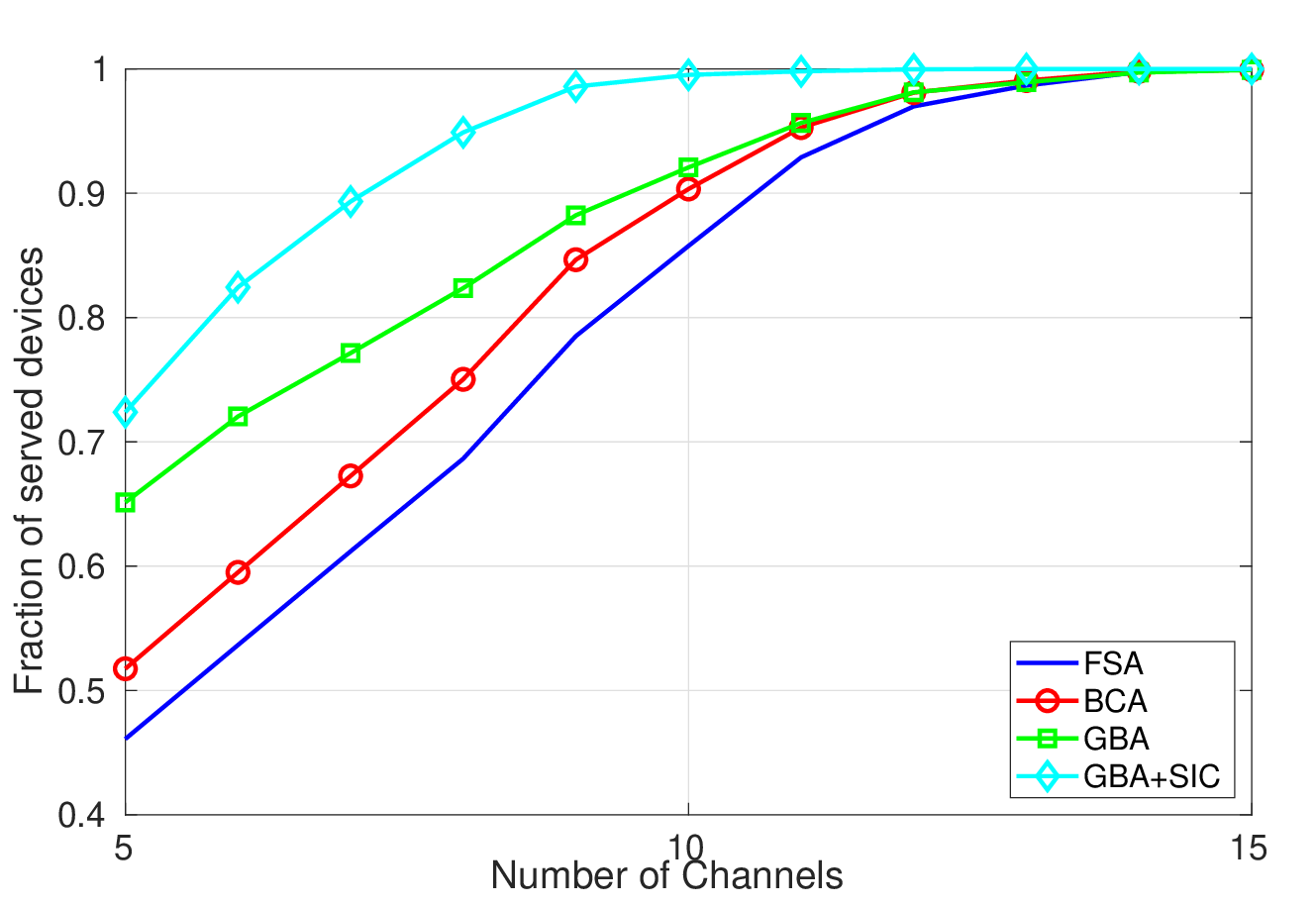}
 \caption{\small Fraction of served devices as a function of the number of channels, with $N=160$ devices.}
 \label{fig:NEW_CON_vs_C}
 \vspace{-0.6cm}
 \end{center}
\end{figure}
We ran a set of simulations to compare the performance of the three algorithms in this scenario, which we report in Figure~\ref{fig:NEW_CON_vs_N} and in Figure~\ref{fig:NEW_CON_vs_C}.
As expected, while the performance of BCA and GBA remains unaltered, the one of FSA highly improves when the fading is constant across the channels. Nonetheless, both BCA and GBA still outperform FSA, due to the fact that they avoid allocating data over channels with high interference whenever it is possible.
In a more general scenario with correlated channels, the performance of FSA lies in between the curve reported here and the one depicted in Section~\ref{sec:resu}, which confirms the better allocation offered by GBA even when channel fading is correlated.

\bibliographystyle{IEEEtran}
\bibliography{IEEEabrv,biblio}

\end{document}